\documentclass[twocolumn]{aastex631} 
\usepackage{amsmath}
\usepackage{color}

\def\gsim{\mathrel{\rlap{\lower 4pt \hbox{\hskip 1pt $\sim$}}\raise 1pt
\hbox {$>$}}}
\def\lsim{\mathrel{\rlap{\lower 4pt \hbox{\hskip 1pt $\sim$}}\raise 1pt
\hbox {$<$}}}
\begin{document}

\title{
Insights into the Properties of Type Ibn/Icn Supernovae and Their Progenitor Channels through X-ray Emission}

\correspondingauthor{Yusuke Inoue}
\email{yusuke@kusastro.kyoto-u.ac.jp}

\author[0009-0004-3148-0462]{Yusuke Inoue}
\author[0000-0003-2611-7269]{Keiichi Maeda}
\affiliation{Department of Astronomy, Kyoto University, Kitashirakawa-Oiwake-cho, Sakyo-ku, Kyoto, 606-8502. Japan; yusuke@kusastro.kyoto-u.ac.jp}

\begin{abstract}

Type Ibn/Icn supernovae (SNe Ibn/Icn), which are characterized by narrow helium or carbon lines originated in hydrogen-poor dense circumstellar medium (CSM), provide new insights into the final evolution of massive stars. While SNe Ibn/Icn are expected to emit strong X-rays through the strong SN-CSM interaction, the X-ray emission modeling effort has been limited so far. In the present study, we provide broad-band X-ray light curve (LC) predictions for SNe Ibn/Icn. We find that the soft X-ray LC provides information about the CSM compositions, while the hard X-ray LC is a robust measure of the CSM density, the explosion energy, and the ejecta mass. In addition, considering the evolution of the ionization state in the unshocked CSM, a bright soft X-ray is expected in the first few days since the explosion, which encourages rapid X-ray follow-up observations as a tool to study the nature of SNe Ibn/Icn. Applying our model to the soft X-ray LCs of SNe Ibn 2006jc and 2022ablq, we derive that the CSM potentially contains a larger fraction of carbon and oxygen for SN 2006jc than 2022ablq, highlighting the power of the soft X-ray modeling to address the nature of the CSM. We also discuss detectability and observational strategy, with which the currently operating telescopes such as NuSTAR and Swift can offer an irreplaceable opportunity to explore the nature of these enigmatic rapid transients and their still-unclarified progenitor channel(s).

\end{abstract}


\section{Introduction} \label{sec:intro}
    Massive stars with the zero-age main-sequence mass ($M_{\mathrm{ZAMS}}$) exceeding $\sim8\ \mathrm{M_\odot}$ trigger terminal explosions called core-collapse supernovae (CCSNe). Most of them show broad lines corresponding to the ejecta velocity of $\sim10^{4}\ \mathrm{km\ s^{-1}}$ in their spectra, while some SNe show narrow lines. For example, Type IIn SNe (SNe IIn) show narrow hydrogen (H) lines \citep{1997ARA&A..35..309F}. The narrow lines are believed to form in the dense circumstellar medium with the velocity of $\sim10^{2}\ \mathrm{km\ s^{-1}}$
    \citep[e.g.,][]{2014ARA&A..52..487S,2017hsn..book..403S}. 
    The collision of SN ejecta with such dense CSM efficiently dissipates the kinetic energy, leading to strong emissions across the multi wavelengths \citep[e.g., optical, radio and X-ray;][]{2007ApJ...659L..13O,2007ApJ...666.1116S,2012ApJ...756..173S,2012ApJ...755..110C,2014ApJ...781...42O,2014ApJ...797..118F}. This picture of the SN-CSM interaction model for SNe IIn has been well established \citep[e.g.,][]{1982ApJ...258..790C, 2003LNP...598..171C, 2017hsn..book..875C}.\par 
    Their CSM distribution reflects the mass-loss history before the SN explosion. For SNe IIn, the optical LC modeling based on the SN-CSM interaction model \citep[e.g.,][]{2013MNRAS.435.1520M, 2014MNRAS.439.2917M} has indicated extreme mass loss rates of $10^{-4}\ \mathrm{{M}_{\odot}}\sim1\ \mathrm{{M}_{\odot}} \mathrm{yr}^{-1}$ within $\sim100$ years toward the explosion. Such LC analyses provide valuable opportunities for revealing stellar evolution and explosion properties.\par
    There are several variants in the family of interacting SNe. 
    SNe Ibn show narrow helium (He) lines, and likely have He-rich CSM; SNe Icn show narrow carbon (C) lines, and likely have C-rich CSM. The widths of these herium and carbon lines correspond to the CSM velocity of $\sim10^{3}\ \mathrm{km\ s^{-1}}$ \citep{2007Natur.447..829P,2016MNRAS.461.3057S,2017MNRAS.471.4381S,2017ApJ...836..158H,2022ApJ...927..180P,2022ApJ...938...73P,2022Natur.601..201G,2023A&A...673A..27N,2023MNRAS.523.2530D,2023ApJ...959L..10P}.
    The progenitors of SNe Ibn/Icn might have undergone a more extreme mass-loss episode than the progenitors of SNe IIn, with the mass loss reaching down to the inner He or even C layer \citep[e.g.,][]{2007Natur.447..829P,2022Natur.601..201G}.\par
    The optical LCs of SNe Ibn/Icn show a rapid decline. It has been interpreted that SNe Ibn/Icn are mainly powered by the SN-CSM interaction with a steep CSM density profile \citep{2016ApJ...824..100M, 2022ApJ...927...25M,2023A&A...673A..27N}. The steep CSM profile translates into the mass-loss rate of the progenitor increasing toward the explosion.
    This may provide a clue to the still-unclarified mechanism(s) of such pre-SN activity \citep[e.g.,][]{2007ApJ...657L.105F, 2020MNRAS.491.6000S,2022ApJ...927...25M,2024A&A...684L..18B}.\par
    However, the optical LC modeling involves relatively large uncertainties \citep[e.g.,][]{2020ApJ...889..170G,2022ApJ...926..125P}, since the optical emission is an outcome of reprocessing the originally emitted X-ray photons to optical photons through various processes such as Compton scattering. The X-ray emission can provide a more direct view than the optical emission, on the properties of the CSM and SN-ejecta; the temperatures behind forward shock (FS) and reverse shock (RS) correspond to the X-ray energy range.\par
    The X-ray observation of SNe Ibn/Icn has been limited due to their rarity and the rapidly-evolving nature. The only two cases with the reported X-ray detection are SNe Ibn 2006jc \citep{2008ApJ...674L..85I} and 2022ablq \citep{2024arXiv240718291P}, and there are a few cases where upper limits are placed, e.g., SN Icn 2019hgp \citep{2022Natur.601..201G}. However, once obtained, the X-ray data provide powerful diagnostics; for SN 2006jc, \citet{2009MNRAS.400..866C} extracted the properties of the CSM through the X-ray LC modeling. The present work aims at promoting/accelerating X-ray observations of SNe Ibn/Icn, by providing the X-ray model that can be directly applied to such data.\par
    This paper is structured as follows. The methods of our X-ray LC modeling are described in Section \ref{sec:method}. The results are presented in Section \ref{sec:Results}. In Section \ref{sec:Application}, we apply our model to SNe 2006jc, 2019hgp and 2022ablq. In Section \ref{sec:photoionization}, additional discussion is given on the evolution of the ionization state in the unshocked CSM. Importance of hard X-ray observations and a caveat in our X-ray LC model are described in Section \ref{sec:Discussion}. Our findings are summarized in Section \ref{sec:summary}.\par

\section{Methods}

\label{sec:method}
    We assume that SNe Ibn/Icn are powered solely by the SN-CSM interaction, and omit possible contribution by radioactive decay \citep[e.g.,][]{2007ApJ...657L.105F, 2022ApJ...927...25M}. Behind the forward shock (FS) and the reverse shock (RS), the gas temperatures immediately after the shock heating are $\sim10^{9}\ \mathrm{K}$ and $10^{7}\ \mathrm{K}$, respectively \citep{2003LNP...598..171C, 2017hsn..book..875C, 2022ApJ...927...25M}. The X-ray emission process behind the FS and the RS is dominated by free-free emission and line emission, respectively. In this section, we describe our methods to calculate the X-ray LCs for SNe Ibn/Icn. 
    \subsection{Hydrodynamics for the SN-CSM interaction} \label{subsec:methodofhydrodynamics}
        Hydrodynamics for the SN-CSM interaction is simulated in 1D spherical symmetric geometry, adopting the open-source code SNEC \citep[The SuperNova Explosion Code;][]{2015ApJ...814...63M}.
        The input parameters are the following; ejecta mass ($M_{\mathrm{ej}}$), ejecta (kinetic) energy ($E_{\mathrm{kin,ej}}$), and the CSM density ($\rho_{\mathrm{CSM}}$). The initial ejecta structure as a function of radius ($r$) is assumed to follow the homologous expansion \citep{2003LNP...598..171C,2017hsn..book..875C}, as represented by a broken power-law function; each segment is described as $\rho_\mathrm{ej}\propto r^{-n}$, where $n=0$ for the inner part and $7$ for the outer part \citep[assuming a compact-star progenitor like a Wolf-Rayet star:][]{2008ApJ...687.1208T}. The radial distribution of the CSM density is assumed to follow a single power-law function as $\rho_\mathrm{CSM}=10^{-14}D'(r/5\times10^{14}\ \mathrm{cm})^{-s}\ \mathrm{g\ cm^{-3}}$ \citep[see][]{2022ApJ...927...25M}. Here, the $D'$ is a constant parameter that controls the density scale. This is the same setup used in the previous studies on the optical LC modeling \citep{2022ApJ...927...25M, 2023A&A...673A..27N}. 
        
        A typical range of the parameters we investigate is the following. For SNe Ibn, $M_{\mathrm{ej}}=2-6\ \mathrm{M_\odot}$, $E_{\mathrm{kin,ej}}\sim10^{51}\ \mathrm{erg}\ (=1\ \mathrm{Bethe \ (B)})$, $D'\sim0.5-5.0$, and $s\sim3$ \citep{2022ApJ...927...25M}. For SNe Icn, the parameters are similar to those of SNe Ibn, while the ejecta energy may be higher \citep[a few B;][]{2023A&A...673A..27N}.
        
        With the setup described above, we first calculate the SN-CSM interaction with the adiabatic hydrodynamic mode of SNEC. The radiative cooling by free-free and line emissions, as well as the energy exchange between electrons and ions, are incorporated as a post-process. As the initial condition of the thermal energy evolution of ions, we adopt the gas properties (e.g., gas density and gas temperature) given by the SNEC simulation. Subsequently, we calculate the thermal energy evolution of ions and electrons as described below. For each mass element, the time evolutions of the thermal energy of electrons and that of ions ($\frac{\mathrm{d}E_{\mathrm{th,e}}}{\mathrm{d}t}, \frac{\mathrm{d}E_{\mathrm{th,ion}}}{\mathrm{d}t}$) are computed as follows: 
        \begin{equation}\label{thermalmodel}
            \frac{\mathrm{d}E_{\mathrm{th,e}}}{\mathrm{d}t}=\frac{\mathrm{d}E_{\mathrm{ad,th,e}}}{\mathrm{d}t} - L_{\mathrm{X}}(t) + P_{\mathrm{e-ion}}\ , 
        \end{equation}
        \begin{equation}\label{ionthermalmodel}
            \frac{\mathrm{d}E_{\mathrm{th,ion}}}{\mathrm{d}t}=\frac{\mathrm{d}E_{\mathrm{ad,th,ion}}}{\mathrm{d}t} - P_{\mathrm{e-ion}}\ , 
        \end{equation}
        where $\frac{\mathrm{d}E_{\mathrm{ad,th,}i}}{\mathrm{d}t}(i=\mathrm{e, ion})$ expresses the time evolution of the thermal energy of electrons or ions in the mass element, only taking into account the contribution from the adiabatic expansion cooling and the shock heating. In computing the X-ray luminosity (i.e., radiative cooling), $L_{\mathrm{X}}$, we take into account free-free emission and line emissions (see Section \ref{subsec:methodofemission} for details). Practically, we place a limit on $L_{\mathrm{X}}$ so that the internal energy does not become negative. The term $P_{\mathrm{e-ion}}$ expresses the energy transfer from ions to electrons. 
        In eq. \ref{ionthermalmodel}, we treat the gas including various ions as a single-temperature component. 
        
        Using eqs. \ref{thermalmodel}, \ref{ionthermalmodel} and the Ideal single-particle Boltzmann Gas EOS with the output from the adiabatic hydrodynamics simulations by SNEC, the electron and ion temperatures ($T_{\mathrm{e}}, T_{\mathrm{ion}}$) in the downstream of a shock wave are calculated. The ratio of specific heats is chosen to be 5/3.\par
        The adiabatic term is expressed as follows:
        \begin{equation}\label{internal_energy_changing}
            \frac{\mathrm{d}E_{\mathrm{ad,th,}i}}{\mathrm{d}t}=
            \begin{cases}
            \frac{\mathrm{d}E_{\mathrm{ad,th}}}{\mathrm{d}t}\frac{n_{i}}{n_{\mathrm{ion}}+n_{\mathrm{e}}}&\text{(shock front)}\\
            \frac{\mathrm{d}E_{\mathrm{ad,th}}}{\mathrm{d}t}\frac{T_{i}}{T}\frac{n_{i}}{n_{\mathrm{ion}}+n_{\mathrm{e}}}&\text{(downstream of the shock)}
            \end{cases}
        \end{equation}
        The gas temperature taken from the adiabatic hydrodynamics simulations by SNEC is denoted by $T$, while the electron and ion temperatures computed through the post process are expressed by $T_{\mathrm{e}}$ and $T_{\mathrm{ion}}$. The $n_{\mathrm{e}}$ and $n_{\mathrm{ion}}$ are the number density of free electrons and that of ions, respectively (see Section \ref{subsec:methodofabsorption} for the treatment of ionization state). 
        The upper term of eq. \ref{internal_energy_changing} includes the contributions from the shock heating and the ion-electron energy exchange, assuming that the energy exchange between electrons and ions is immediately completed at the shock front. This is a rational assumption since the relaxation timescale between electrons and ions is much shorter than the dynamical timescale under the situation considered here \citep[see][]{2003LNP...598..171C,2006ApJ...651..381C,2022ApJ...927...25M}. The factor $\frac{n_{i}}{n_{\mathrm{ion}}+n_{\mathrm{e}}}$ takes into account the contribution of free electrons to the time evolution of thermal energy ($\frac{\mathrm{d}E_{\mathrm{ad,th}}}{\mathrm{d}t}$) in the adiabatic hydrodynamics simulation by SNEC. The lower term of eq. \ref{internal_energy_changing} describes the adiabatic expansion cooling of electrons. The factor $\frac{T_{i}}{T}$ is introduced to correct for the effect of the adiabatic cooling after taking into account the radiative cooling of electrons.\par
        The electron-ion energy-exchange rate, $P_{\mathrm{e-ion}}$, is expressed as follows (see eq. 5 in \citet{1977PASJ...29..813I} and eq. 26 in \citet{2006ApJ...651..381C}):
        \begin{equation}\label{exchangeionele}
            P_{\mathrm{e-ion}}=\frac{3}{2}\frac{\bar{Z}}{4}\frac{\ln\Lambda}{4.2\times 10^{-22}} k_{\mathrm{B}}(T_{\mathrm{ion}}-T_{\mathrm{e}})n_{\mathrm{e}}\rho T_{\mathrm{e}}^{-1.5}, 
        \end{equation}
        where $\bar{Z}$, $\ln\Lambda\approx 30$, $k_{\mathrm{B}}$, and $\rho$ are the average charge of the ions, the Coulomb logarithm, the Boltzmann's constant, and the mass density of the gas given by the SNEC simulation. In our reference model (Section \ref{sec:Results}), $\bar{Z}\approx3$ corresponding to (He, C, O)=(0.5, 0.25, 0.25) in the mass fraction\footnote{Throughout this paper, the compositions are expressed in the mass fractions unless mentioned.}. At the shock front, we set as $P_{\mathrm{e-ion}}=0$ because eq. \ref{internal_energy_changing} already includes the effect of $P_{\mathrm{e-ion}}$ there.\par
        After the calculation of the thermal energy evolution, the X-ray luminosity at a given energy band is calculated considering absorption processes such as Compton scattering and photoelectric absorption (see Section \ref{subsec:methodofabsorption}). 
        We note that our model for the internal energy evolution does not include the feedback effect of the radiation loss on hydrodynamics, but we confirm that this treatment represents a good first approximation (see Appendix \ref{subsec:radiationhydro}).

    \subsection{X-ray emission process}\label{subsec:methodofemission}
        The X-ray emission from SNe Ibn/Icn is dominated by that arising from the FS for the following reasons. Because of the dense CSM, the RS is ineffective in producing X-ray emission at least until a few 100 days; the X-ray emission from the RS is mainly in soft X-ray band ($\sim0.1-10\ \mathrm{keV}$), and initially the soft X-rays are almost entirely absorbed by photoelectric absorption in the cooled shocked ejecta behind the RS (see Sections \ref{subsec:methodofabsorption} and \ref{subsec:softx}). Also, the cooling is so effective behind the RS, and the RS region is rapidly cooled down; thus little X-ray emission is expected from the RS. The cooling in the FS region is much slower than the RS region due to much higher temperature; when the time evolution of the thermal energy (eq. \ref{thermalmodel}) is calculated, the line emission is taken into account using the fitting formula for the cooling function presented by \citet{2003LNP...598..171C, 2017hsn..book..875C}, but it is inefficient at $\sim10^{9}\ \mathrm{K}$. Therefore, free-free emission from the FS region is dominant in the entire X-ray bands covering the soft ($<10\ \mathrm{keV}$) and hard ($>10\ \mathrm{keV}$) X-rays, and we compute the X-ray spectral energy distribution (SED) under this assumption. 
        
        The emissivity of free-free emission is expressed as follows \citep{1979rpa..book.....R}:
        \begin{equation}\label{ffemi}
            \varepsilon^{\mathrm{ff}}_{\nu}=6.84\times 10^{-38} g_{\mathrm{ff}}T_{\mathrm{e}}^{-0.5} n_{\mathrm{e}} \sum_{j} Z_{j}^{2}n_{j} \exp\left(-\frac{E_{\mathrm{ph}}}{\mathrm{k_{B}}T_{\mathrm{e}}}\right),
        \end{equation}
        where $g_{\mathrm{ff}}$, $Z_{j}$, $n_{j}$, and $E_{\mathrm{ph}}$ are the gaunt factor, the atomic number of element $j$, the number density of element $j$, and the photon energy.
        $n_{\mathrm{e}}$ and $n_{\mathrm{j}}$ are calculated using the mass density of the gas (see Section \ref{subsec:methodofabsorption} for the treatment of ionization state).
        The gaunt factor $g_{\mathrm{ff}}$ is fixed to be 1 for simplicity.
        
    \subsection{X-ray absorption processes}\label{subsec:methodofabsorption}
        From optical spectra of SNe Ibn/Icn, it is believed that they have H-poor CSM \citep[e.g.,][]{2007Natur.447..829P,2022Natur.601..201G}. We assume that the CSM is composed of elements heavier than H such as He, C, oxygen (O), Neon (Ne), and Magnesium (Mg), and set their mass fractions guided by previous study and stellar evolution simulations \citep{1988PhR...163...13N, 2008ApJ...687.1208T}. For example, \citet{2009MNRAS.400..866C} and \citet{2008MNRAS.389..141M} have chosen (He, C, O)$\sim$(0.73, 0.14, 0.13) as the mass fractions. In the present study, we investigate X-ray emission from SNe Ibn and Icn. Therefore, we choose the wide range of CSM compositions (see Sections \ref{subsec:softx} and \ref{subsubsec:SN2019hgp}). 
        
        The ionization state in the shocked region is modeled as follows, assuming that the collisional ionization is the main ionization process \citep[see][for details]{1969ApJ...157.1157C}. We treat all atoms as effectively neutral at $T_{\mathrm{e}}<10^{4}\ \mathrm{K}$; at $T_{\mathrm{e}}=10^{4}-10^{6}\ \mathrm{K}$, He is fully ionized, and the others are He-like ions; at $T_{\mathrm{e}}=10^{6}-10^{7}\ \mathrm{K}$, He, C, and O are fully ionized, and the others are He-like ions; at $T_{\mathrm{e}}>10^{7}\ \mathrm{K}$, all atoms are fully ionized. The results of LCs are not sensitive to this assumption on the ionization state in the shocked region. This is because the temperature in the region behind the FS is $\sim10^{9}\ \mathrm{K}$, thus all atoms (even iron (Fe)) can be assumed to be fully ionized. On the other hand, we assume that atoms in the unshocked CSM are (effectively) neutral in Sections \ref{sec:Results} and \ref{sec:Application} (see Section \ref{sec:photoionization} for the effect of photoionization in the unshocked CSM).
 
        We consider Compton scattering and photoelectric absorption as absorption processes for X-rays. We use the angle-averaged Klein-Nishina's formula \citep{1979rpa..book.....R} for Compton scattering. Photoelectric absorption is calculated using open database xlaylib \citep{2004AcSpB..59.1725B, 2011AcSpB..66..776S}. The xlaylib provides opacites of photoelectric absorption for neutral and ground state atoms. We treat ions which are not fully ionized as neutral atoms in the calculation of absorption. The resulting LCs are not sensitive to this assumption, since the photoelectric absorption is mainly contributed by electrons in the K-shell of ions.\par
        The total opacity of photoelectric absorptions by all ions ($\kappa_{\mathrm{pe,total}}(E_{\mathrm{ph}})$), for photons with $E_{\mathrm{ph}}$, is described as follows:
        \begin{equation}\label{tauphotoele}
            \kappa_{\mathrm{pe,total}}(E_{\mathrm{ph}})=\sum_{j} Y_{j} \kappa_{\mathrm{pe},j}(E_{\mathrm{ph}}),
        \end{equation}
        where $Y_{j}$ is the mass fraction of element $j$. For fully ionized elements, $Y_{j}$ is set to be $0$. In the above equation, $\kappa_{\mathrm{pe},j}(E_{\mathrm{ph}})$ is the opacity of photoelectric absorption by neutral atom $j$ for photons with $E_{\mathrm{ph}}$, and this is given by xraylib. 
        
        The total opacity at $E_{\mathrm{ph}}$ in each mass element is then expressed as follows:
        \begin{equation}\label{tautotal}
            \kappa(E_{\mathrm{ph}})=\kappa_{\mathrm{pe,total}}(E_{\mathrm{ph}})+\kappa_{\mathrm{comp}}(E_{\mathrm{ph}}),
        \end{equation}
        where $\kappa_{\mathrm{comp}}(E_{\mathrm{ph}})$ is the opacity of Compton scattering. Compton scattering dominates the opacity in the FS region, while photoelectric absorption is the most important in the RS region. Indeed, the unshocked CSM is the main contributor to the photoelectric absorption. Therefore, we consider Compton scattering only in the FS region.   \par
        For X-rays emitted from each mass element, the observed luminosity ($L_{\mathrm{X,obs}}(E_{\mathrm{ph}})$) is expressed as follows:
        \begin{equation}\label{obsxray}
            L_{\mathrm{X,obs}}(E_{\mathrm{ph}})=0.5L_{\mathrm{X}}(E_{\mathrm{ph}})\exp(-\tau(E_{\mathrm{ph}})),
        \end{equation}
        \begin{equation}\label{obsxray_2}
            \tau(E_{\mathrm{ph}})=\int_{r_{\mathrm{emi}}}^{\infty} \kappa(E_{\mathrm{ph}})\rho \mathrm{d}r.
        \end{equation}
        Here, the $r_{\mathrm{emi}}$ is where the radiation $L_{\mathrm{X}}(E_{\mathrm{ph}})$ is emitted. The factor of 0.5 in eq. \ref{obsxray} comes from our assumption that half of isotropically emitted X-rays goes outward following eq. \ref{obsxray} while the other half goes inward, a frequently-adopted assumption in such analyses \citep[e.g.,][]{2003LNP...598..171C,2022ApJ...927...25M}. The X-rays that go inward are converted into low energy photons (e.g., optical or UV) in the SN ejecta. \par

    \begin{figure}[t]
    \centering
    \includegraphics[scale=0.6]{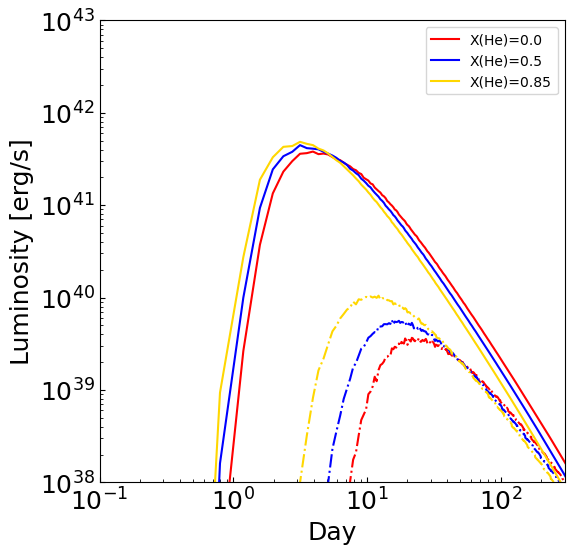}
    \caption{The dependence of the synthetic hard (10-40 keV; solid) and soft (0.2-10 keV; dash-dotted) X-ray LCs on the CSM composition. Shown here are the models with X(He)=0 (red), 0.5 (blue), and 0.85 (yellow). 
    \label{SNeIbn_x-ray_abundance}}
    \end{figure}

    \begin{figure}[t]
    \centering
    \includegraphics[scale=0.6]{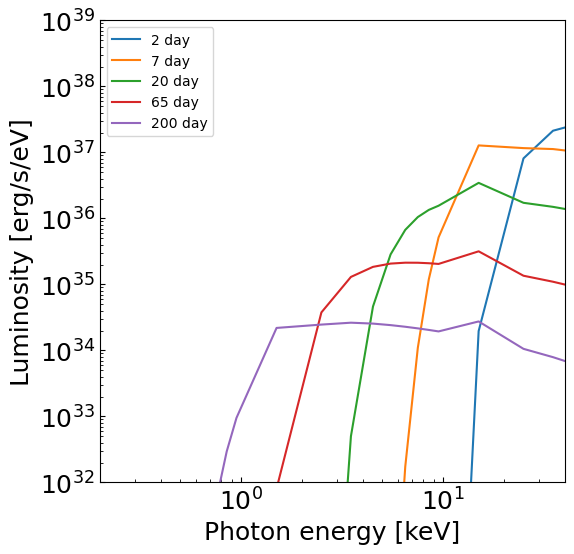}
    \caption{The SED evolution for the model with (He, C, O)=(0.5, 0.25, 0.25). The blue, orange, green, red, and violet solid lines show the SED at 2, 7, 20, 65, 200 days since the explosion, respectively.
    \label{D=1.6,SED}}
    \end{figure}

    \begin{figure*}[t]
    \begin{center}
        \begin{minipage}[b]{0.45\linewidth}
            \centering
            \includegraphics[keepaspectratio, scale=0.35]{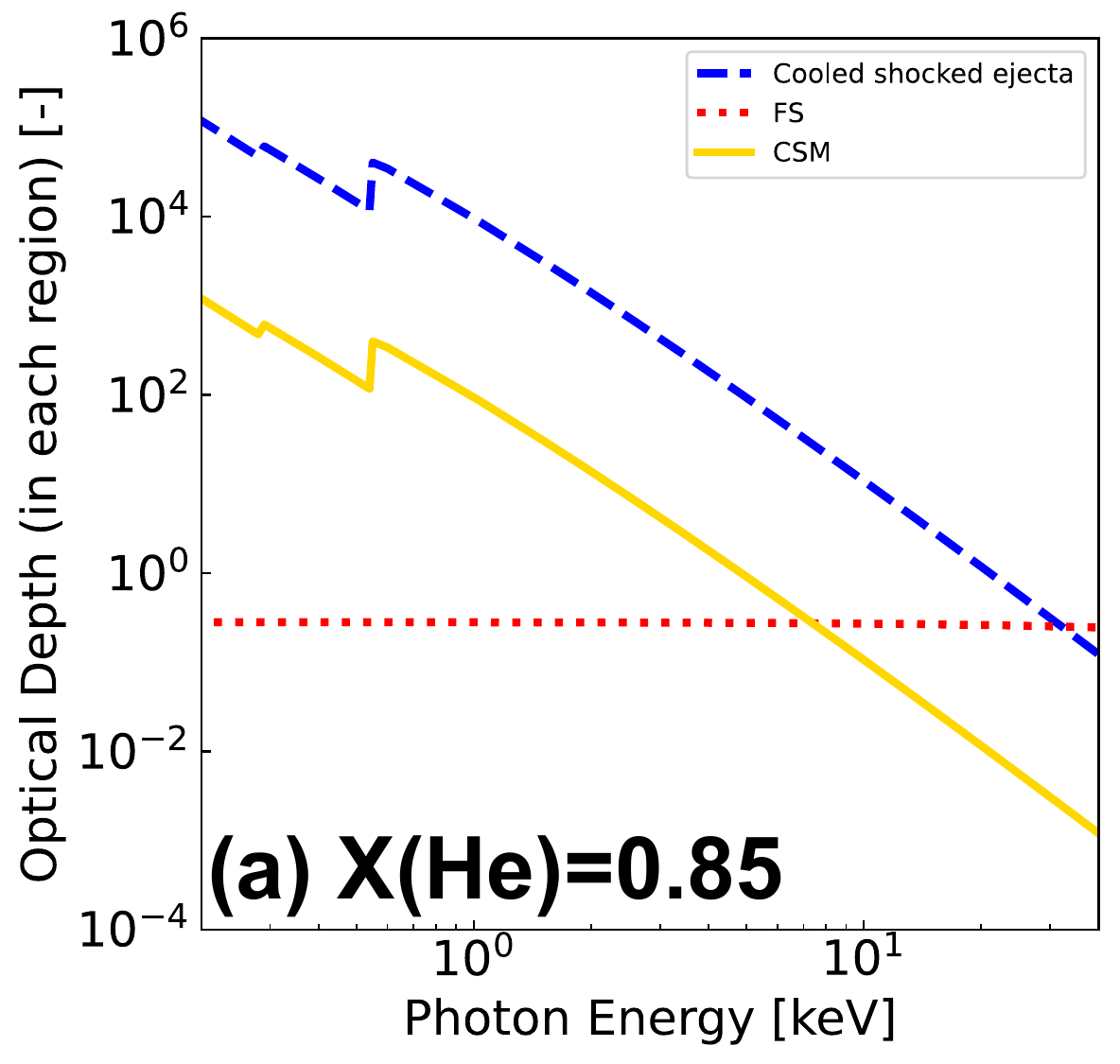}
        \end{minipage}
        \hspace{0.1\columnwidth}
        \begin{minipage}[b]{0.45\linewidth}
            \centering
            \includegraphics[keepaspectratio, scale=0.35]{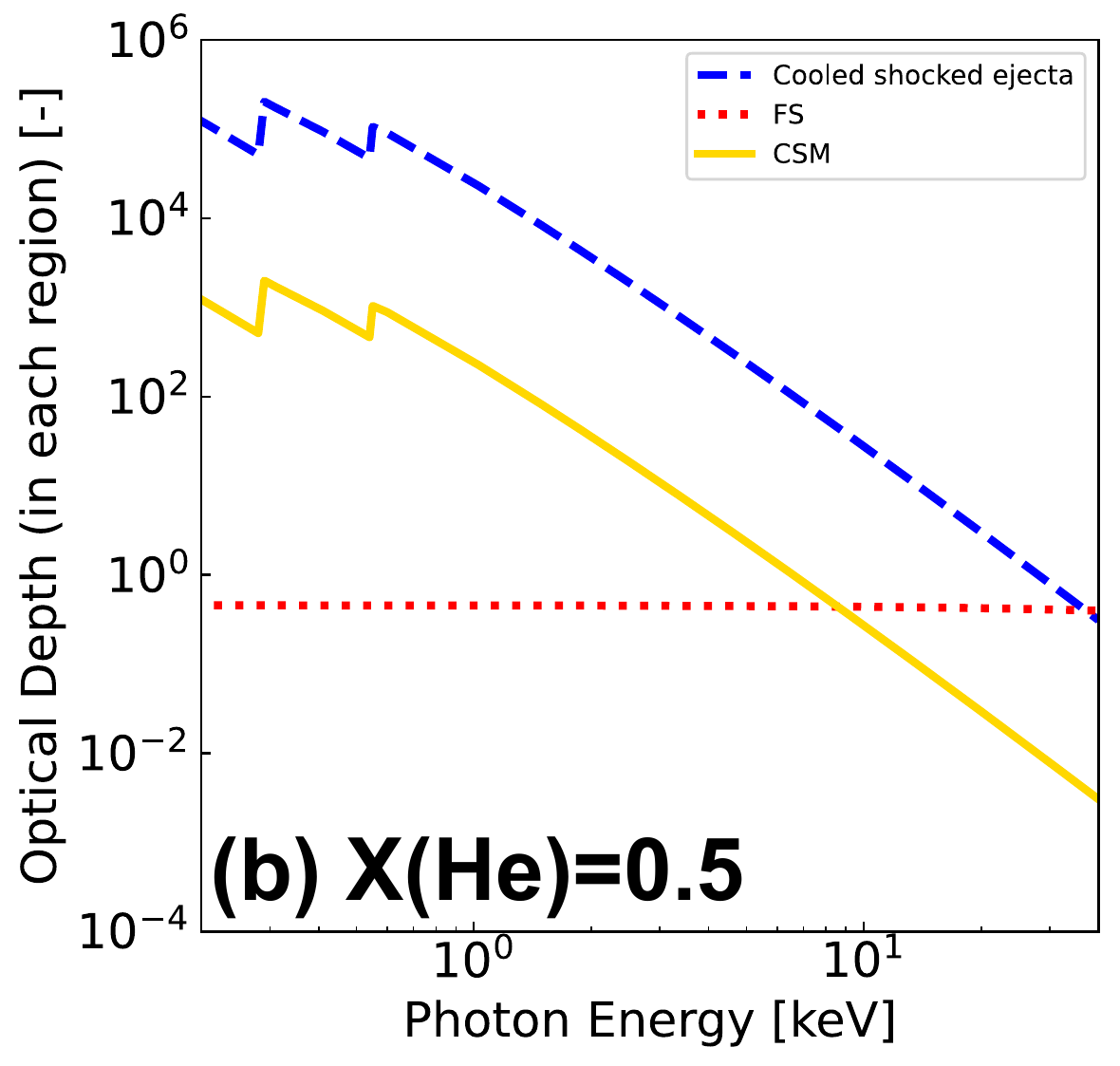}
        \end{minipage}
        \\
        \begin{minipage}[b]{0.45\linewidth}
            \centering
            \includegraphics[keepaspectratio, scale=0.35]{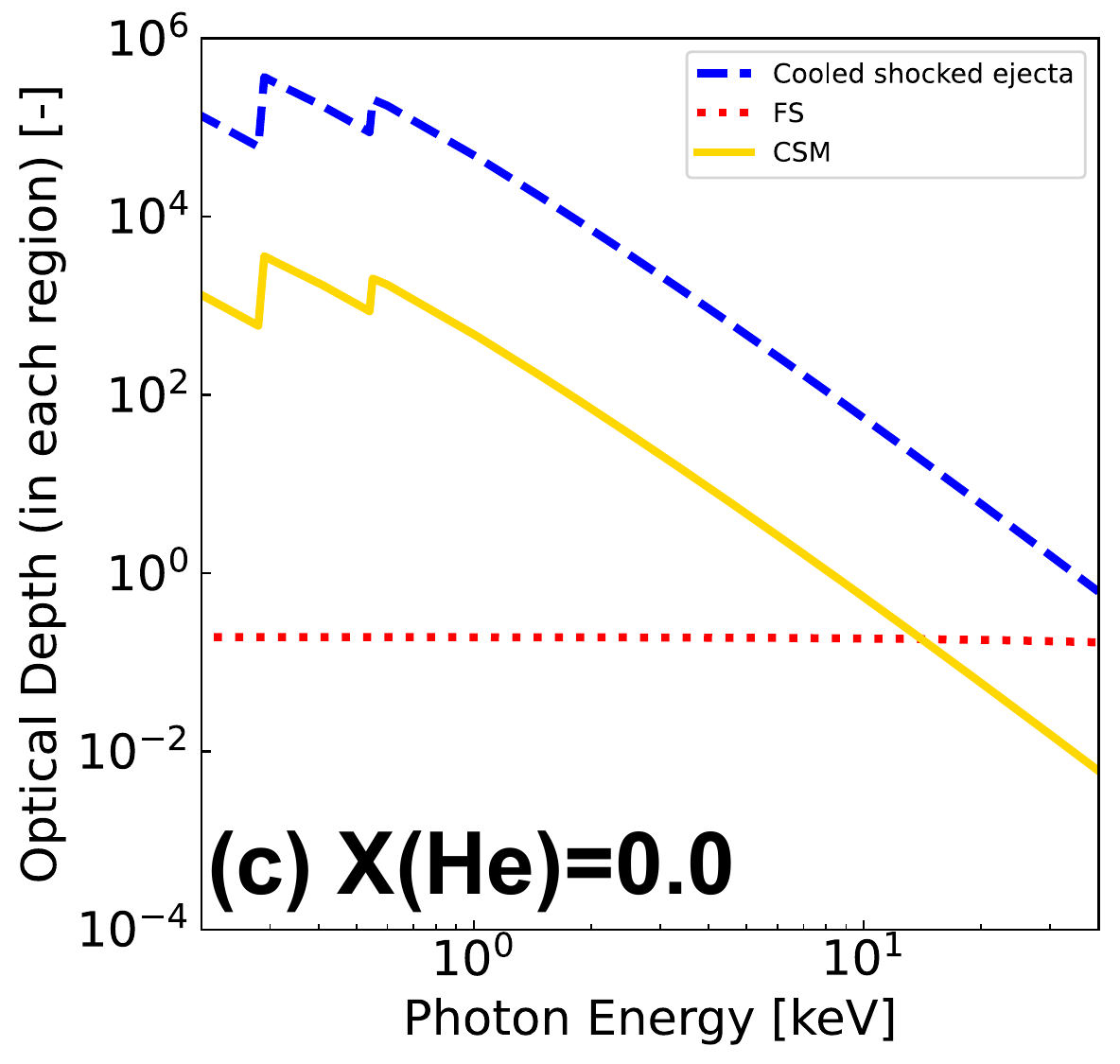}
        \end{minipage}
        \caption{The dependence of the optical depth on the CSM composition; (a) X (He)=0.85, (b) X (He)=0.5, and (c) X (He)=0. Shown here are the optical depth of Compton scattering within the region shocked by the FS (dotted), that of photoelectric absorption within the unshocked CSM (solid), and that of photoelectric absorption within the cooled shocked ejecta (dashed)}.\label{SNeIbn,opticaldepth,abundance}
    \end{center}
    \end{figure*}

\section{General properties of X-ray Light-curves with neutral unshocked CSM} \label{sec:Results}
    In this Section, we examine parameter dependence of our X-ray LC model, under the assumption that the unshocked CSM is effectively neutral (i.e., not fully ionized).
    We adopt $M_\mathrm{ej}=2\ \mathrm{M_\odot}$, $E_\mathrm{kin,ej}=1\ \mathrm{B}$, $V_\mathrm{CSM}=1000\ \mathrm{km\ s^{-1}}$, $s=3$, $D'=1.6$, and (He, C, O)=(0.5, 0.25, 0.25) for the CSM composition, unless specifically mentioned. The CSM velocity ($V_\mathrm{CSM}$) is assumed to be $1000\ \mathrm{km\ s^{-1}}$, as is typically measured from the narrow lines of SNe Ibn/Icn. 
    
    \subsection{Soft X-ray properties}\label{subsec:softx}
        The property of the soft X-ray strongly reflects the CSM composition through photoelectric absorption \citep[$\kappa_{\mathrm{pe}}\propto E_\mathrm{ph}^{-3.5}$ for the K-shell approximation;][]{2011hea..book.....L}.
        Here, we adopt He, C and O as representative elements in the CSM. A model with (He, C, O)=(0.5, 0.25, 0.25) is taken as a reference model, while additional models are examined with (He, C, O)=(0.85, 0.03, 0.12) (helium-envelope composition) and (He, C, O)=(0.0, 0.5, 0.5) (carbon-layer composition).\par
        Figure $\ref{SNeIbn_x-ray_abundance}$ shows the synthetic X-ray LCs. For the lower fraction of the heavy elements, the X-ray LC peaks earlier with a higher peak luminosity because of the metallicity dependence of $\kappa_{\mathrm{pe},j}$ \citep[$\kappa_{\mathrm{pe},j}\propto Z_{j}^{5}$ for the K-shell approximation;][]{2011hea..book.....L}. \par
        Figure \ref{D=1.6,SED} shows the evolution of the Spectral Energy Distribution (SED) of the reference model with X(He)=0.5. At a higher energy, the system becomes optically thin more rapidly. The soft X-ray reaches to a fainter peak luminosity than the hard X-ray. These are general trends for SNe Ibn/Icn, as arising from a combination of the stronger photoelectric absorption at a lower X-ray energy range (see Figure \ref{SNeIbn,opticaldepth,abundance}) and the decreasing SN-CSM interaction power. We note that the contribution of X-ray emission from RS can be ignored because the optical depth in the cooled shocked ejecta behind the RS is much larger than those of the FS and CSM.\par

    \subsection{Hard X-ray properties}\label{subsubsec:ejecta}
        The CSM composition dependence of the hard X-ray LC is weaker than that of the soft X-ray LC (see Figure \ref{SNeIbn_x-ray_abundance}).
        Figure $\ref{SNeIbn,x-ray,Ejecta}$ shows the synthetic X-ray LCs with various ejecta properties ($E_\mathrm{kin,ej}$ and $M_\mathrm{ej}$). With a higher ejecta velocity, the X-ray LC peaks earlier at a higher peak luminosity. This is because the system, including both the shocked and unshocked CSMs, rapidly becomes optically thin for a faster shock wave.\par
        Figure $\ref{SNeIbn,x-ray,Ddash}$ shows the synthetic X-ray LCs with various CSM densities ($D'$). For a higher CSM density, the hard X-ray emerges later and peaks at a higher luminosity; this is a result of the increasing optical depth ($\tau \propto \rho$) and emissivity ($\varepsilon^{\mathrm{ff}}_{\nu}\propto \rho^{2}$) for a higher density. We note that the rising phase of the hard X-ray LCs is not smooth, but this is a numerical artifact; our shortest time step in the LC simulations is $\sim 10^{-2}-10^{-1}$ days.\par
        The hard X-ray LC generally peaks earlier than the soft X-ray LC because the optical depth for the hard X-rays is smaller than that for the soft X-rays. Therefore, there is a period when the rising phase of the soft X-ray LC and the decay phase of the hard X-ray overlap. The decay phase of the hard X-ray LC provides a robust measure of the CSM density irrespective of the evolution in the soft X-ray LC. With the CSM density thus anchored by the hard X-ray properties, the analysis of the rising phase of the soft X-ray LC can then provide the information on the CSM composition separately, i.e., the degeneracy between the CSM density and composition could be solved by these properties (see also Section $\ref{subsec:obspossivility}$).\par

            \begin{figure}[t] 
            \centering
            \includegraphics[keepaspectratio, scale=0.6]{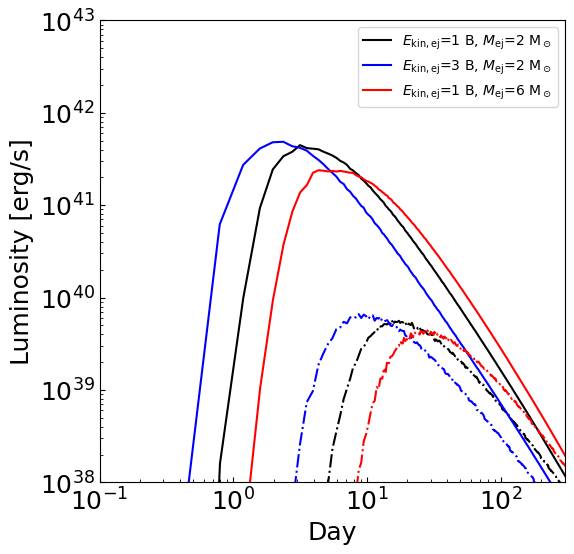}
            \caption{The dependence of the synthetic hard (10-40 keV; solid) and soft (0.2-10 keV; dash-dotted) X-ray LCs on the ejecta properties. Shown here are the models with $E_{\mathrm{kin,ej}}$=1 B and $M_{\mathrm{ej}}$=2 $\mathrm{M_\odot}$ (black), 3 B and 2 $\mathrm{M_\odot}$ (blue), and 1 B and 6 $\mathrm{M_\odot}$ (red).
            \label{SNeIbn,x-ray,Ejecta}}
            \end{figure}
        
            \begin{figure}[t]
            \centering
            \includegraphics[keepaspectratio, scale=0.6]{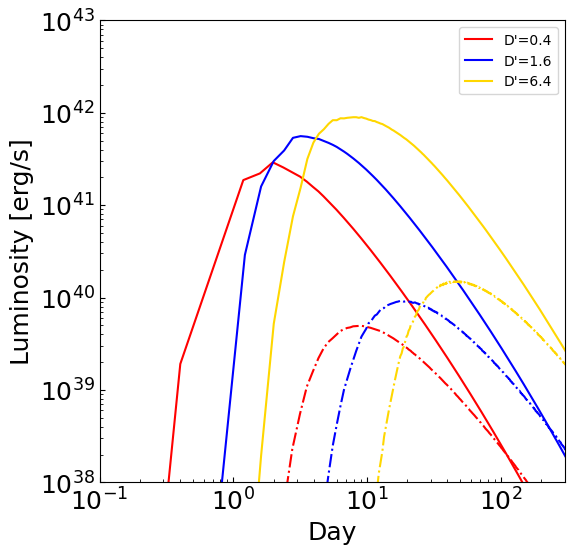}
            \caption{The dependence of the synthetic hard (10-40 keV; solid) and soft (0.2-10 keV; dash-dotted) X-ray LCs on the CSM density. Shown here are the models with $D'=0.4$ (red), $1.6$ (blue), and $6.4$ (yellow). 
            \label{SNeIbn,x-ray,Ddash}}
            \end{figure}

\section{Application to individual objects} \label{sec:Application}
    \subsection{$\mathrm{SN\ 2006jc}$}\label{subsec:2006jc}
        Figure \ref{SN2006jc,xray} shows the synthetic X-ray LCs (with effectively neutral CSM; see Section \ref{sec:photoionization}) as compared to the data of SN 2006jc. The physical parameters adopted here are $M_\mathrm{ej}=6\ \mathrm{M_\odot}$, $E_\mathrm{kin,ej}=0.8\ \mathrm{B}$, $V_\mathrm{CSM}=3000\ \mathrm{km\ s^{-1}}$, $s=3$ and $D'=4.0$. The CSM velocity corresponds to the typical width of the helium emission lines seen in SN 2006jc \citep{2007ApJ...657L.105F, 2007Natur.447..829P, 2008MNRAS.389..113P}. As a reference model, we adopt (He, C, O)=(0.73, 0.14, 0.13) in the mass fractions; this represents the composition in the helium layer and was adopted in the previous studies \citep{2009MNRAS.400..866C, 2008MNRAS.389..141M}. Additionally, we examine the case with (He, C, O)=(0.4, 0.3, 0.3), corresponding to the deeper part of the progenitor toward the carbon layer \citep[e.g.,][]{1988PhR...163...13N, 2008ApJ...687.1208T}. We set the number fraction of carbon larger than that of oxygen in our models; the infrared (IR) spectral evolution of SN 2006jc suggested that carbon grains were formed in the shocked CSM \citep[e.g.,][]{2008ApJ...687.1208T,2009ApJ...692..546S} which requires a larger number density of carbon than oxygen \citep{2008ApJ...684.1343N}.\par
        The model with (He, C, O)=(0.4, 0.3, 0.3) explains the rising phase of SN 2006jc better than the He-rich composition, indicating a nearly complete stripping of the He envelope (see Appendix \ref{subsec:comperror} about robustness in deriving the CSM composition). However, we refrain from deriving a strong conclusion from this result since it is sensitive to the assumed CSM density; our model is based on the optical LC modeling by \citet{2022ApJ...927...25M}, but the optical LC modeling may involve several uncertainties and possible systematic errors. Rather, we emphasize that the robustly-determined CSM density profile is needed to derived the CSM abundance. This is where the hard X-ray observation can play a key role (see Sections \ref{subsubsec:ejecta} and \ref{subsec:obspossivility}).\par
        In this section, photoionization in the unshocked CSM is not considered, and it is assumed that the CSM is not fully ionized. We will show in Section \ref{sec:photoionization} that this assumption is appropriate at $\gsim1$ days since the explosion for SN 2006jc. Therefore, the parameters derived in this section would not be affected by the treatment of the ionization status of the unshocked CSM.

        \begin{figure}[t]
        \centering
        \includegraphics[scale=0.6]{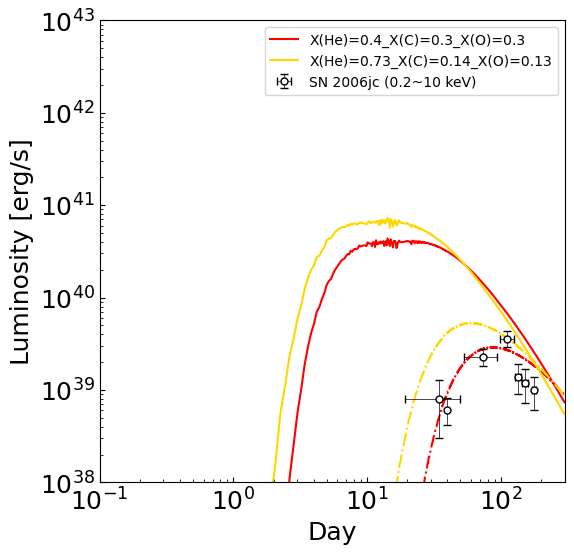}
        \caption{The synthetic hard (10-40 keV; solid) and soft (0.2-10 keV; dash-dotted) X-ray LCs for SN 2006jc with the (effectively) neutral unshocked CSM, as compared to the X-ray data of SN 2006jc. Shown here are the models with (He, C, O)=(0.4, 0.3, 0.3) (red) and (0.73, 0.14, 0.13) (yellow). The data points are from \citet{2008ApJ...674L..85I}. 
        \label{SN2006jc,xray}}
        \end{figure}
        
        \begin{figure}[t]
        \centering
        \includegraphics[keepaspectratio, scale=0.6]{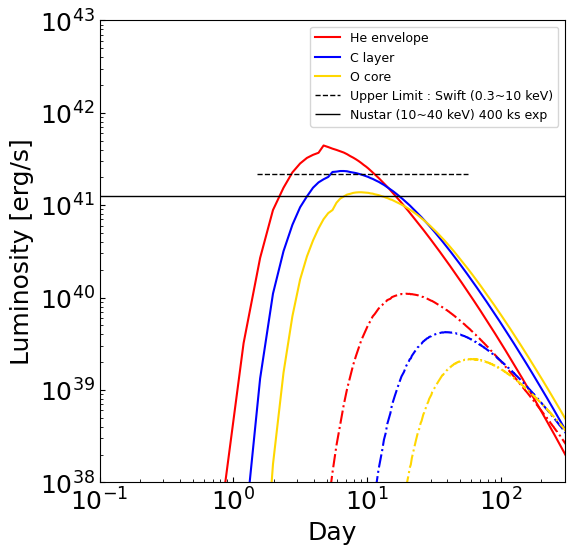}
        \caption{The hard (10-40 keV; solid) and soft (0.2-10 keV; dash-dotted) synthetic X-ray LCs for SN 2019hgp, with different CSM compositions. Shown here are the models with He-envelope compositions (red), the C-layer compositions (blue), and the O-core compositions (yellow). The black-dashed horizontal line represents the soft X-ray upper limit (3$\sigma$) taken from \citet{2022Natur.601..201G}. The black-solid horizontal line shows the 3$\sigma$ sensitivity of NuSTAR with 400 ks exposures.
        \label{SNeIcn,x-ray}}
        \end{figure}

    \subsection{$\mathrm{SN\ 2019hgp}$}\label{subsubsec:SN2019hgp}
        SN 2019hgp is one of the first type Icn supernovae reported in the literature \citep{2021TNSAN..76....1G, 2022Natur.601..201G}. SN 2019hgp was observed with Swift but not detected in 0.3-10 keV during $\sim$1.5-57 days since the explosion. We adopt the upper limits during the entire campaign with Swift reported by \citet{2022Natur.601..201G}. \par
        For SN 2019hgp, we adopt the physical parameters from \citet{2023A&A...673A..27N} based on the optical LC modeling; $M_\mathrm{ej}=3\ \mathrm{M_\odot}$, $E_\mathrm{kin,ej}=2.5\ \mathrm{B}$, $s=2.9$, and $D'=2.3$.  
        In the optical spectra, SNe Icn show lines from elements heavier than SNe Ibn, indicating He-less composition than in SNe Ibn. However, given the difficulty in deriving the He abundance in the optical spectra \citep[e.g.,][]{2010MNRAS.405.2141D,2022A&A...658A.130D} and the interest in constraining the CSM composition independently from the optical spectra, we explore a range of the CSM composition; (He, C, O)=(0.85, 0.03, 0.12) (He-layer) and (0.2, 0.4, 0.4) (C-layer), as well as (C, O, Ne, Mg)=(0.05, 0.65, 0.25, 0.05) (O-core). \par
        Figure $\ref{SNeIcn,x-ray}$ shows that all the models, including the extremely He-rich composition, are consistent with the upper limits of the soft X-ray luminosities of SN 2019hgp; unfortunately the soft X-ray upper limits are more than an order-of-magnitude shallower than the required sensitivity to place a meaningful constraint on the CSM composition. Our model on the other hand suggests that the hard X-rays from SN 2019hgp could probably have been detected by NuSTAR with 400 ks exposure within $\sim5-10$ days; we encourage a prompt harx X-ray observation for a next nearby SN Icn.\par

    \subsection{$\mathrm{SN\ 2022ablq}$}\label{subsec:2022ablq}
        Figure \ref{SN2022ablq,xray} shows the synthetic X-ray LCs (with effectively neutral CSM; see Section \ref{sec:photoionization}) as compared to the data of SN 2022ablq taken from \citet{2024arXiv240718291P}. The physical parameters adopted here are $M_\mathrm{ej}=6\ \mathrm{M_\odot}$, $E_\mathrm{kin,ej}=0.6\ \mathrm{B}$, $V_\mathrm{CSM}=1500\ \mathrm{km\ s^{-1}}$, $s=3$, and $D'=4.0$. We adopt the CSM velocity corresponding to the typical width of the helium emission lines seen in SNe Ibn \citep[e.g.,][]{2017ApJ...836..158H}, following \citet{2024arXiv240718291P}. The adopted CSM density distribution, $s=3$, is a typical value for SNe Ibn \citep{2022ApJ...927...25M}, noting that the optical LC of SN 2022ablq is also typical \citep{2024arXiv240718291P}. We adopt (He, C, O)=(0.95, 0.025, 0.025) corresponding to the helium layer \citep[e.g.,][]{1988PhR...163...13N, 2008ApJ...687.1208T}.\par
        The synthetic soft X-ray LC can explain the observed X-ray LC of SN 2022ablq (see Appendix \ref{subsec:comperror} about the robustness in deriving the CSM composition). This indicates that the CSM around SN 2022ablq contains a smaller fraction of C and O than in SN 2006jc's CSM; the shallower and outer region of the progenitor has been stripped for SN 2022ablq than SN 2006jc. To further straighten this conclusion, we emphasize again that the hard X-ray observation is important for robustly determining the CSM density profile, which is needed to derive the CSM abundance accurately by using the rising phase of the soft X-ray LC (see Sections \ref{subsubsec:ejecta} and \ref{subsec:obspossivility}).\par

        \begin{figure}[t]
        \centering
        \includegraphics[scale=0.6]{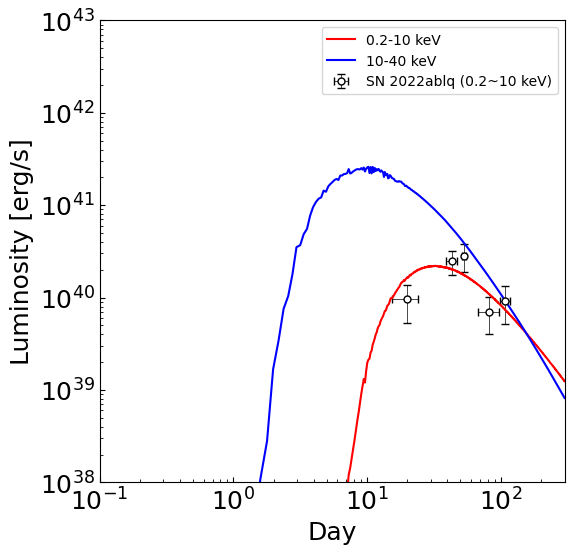}
        \caption{The synthetic hard (10-40 keV; solid-blue) and soft (0.2-10 keV; solid-red) X-ray LCs for SN 2022ablq with the (effectively) neutral unshocked CSM, as compared to the X-ray data of SN 2022ablq. Shown here is the model with (He, C, O)=(0.95, 0.025, 0.025). The data points are from \citet{2024arXiv240718291P}. 
        \label{SN2022ablq,xray}}
        \end{figure}

    \section{The effect of photoionization in the unshocked CSM}\label{sec:photoionization}        
        \begin{figure*}[t]  
            \centering
            \includegraphics[keepaspectratio, scale=0.6]{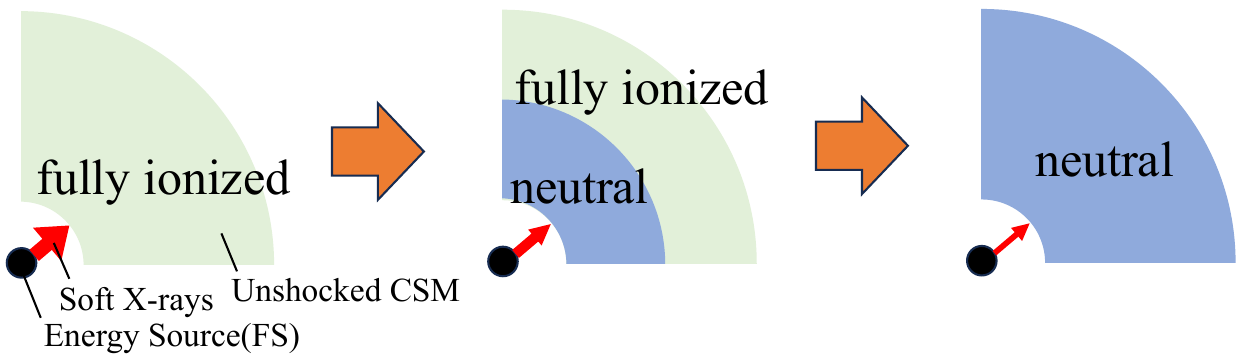}
            \caption{Schematic illustration of the photoionization change, for $\rho\propto r^{-3}$. \label{Photoionization_Changing_Overview}}
        \end{figure*}

        \begin{figure}[t]  
            \centering
            \includegraphics[keepaspectratio, scale=0.6]{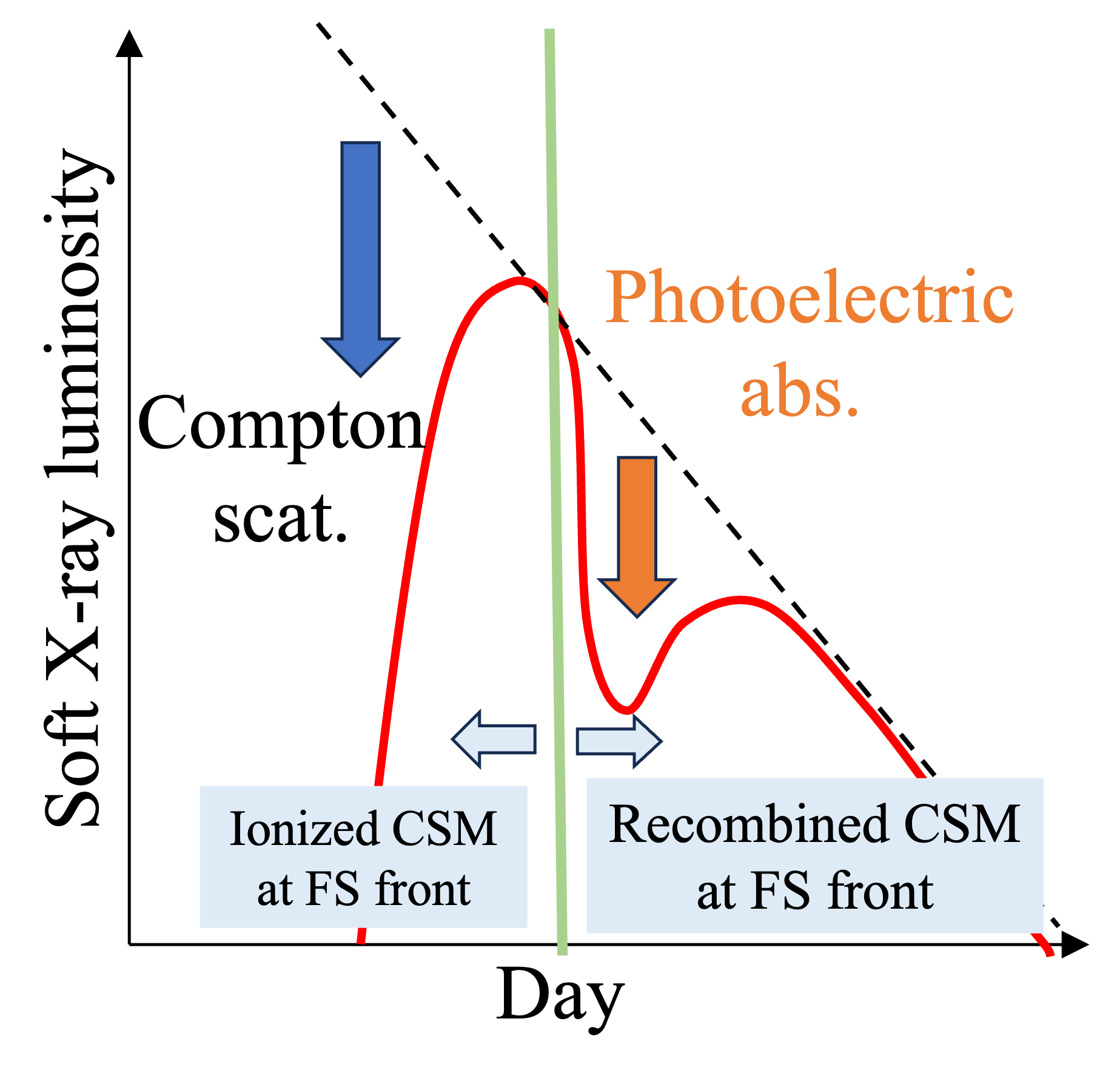}
            \caption{Schematic illustration of the double-peak soft X-ray LC.  The black-dashed line shows the unabsorbed soft X-ray LC. The red-solid line shows the absorbed (observed) soft X-ray LC.\label{DoublePeakLC_Overview}}
        \end{figure}

        \subsection{Calculations of the ionization status}\label{subsec:photoionization}
            Ions in the unshocked CSM absorb soft X-rays mainly by photoelectric absorption (see Figure \ref{SNeIbn,opticaldepth,abundance}) and lose electrons in the K or L-shell. Our numerical simulations in Sections \ref{sec:Results} and \ref{sec:Application} assume that ions in the unshocked CSM are effectively neutral for X-rays, i.e., electrons in the K or L-shell are intact. In Section \ref{sec:photoionization}, we discuss the ionization state of the unshocked CSM, based on the ionization parameter $\xi$ \citep{1969ApJ...156..943T,1976ApJ...206..847H,1982ApJS...50..263K,2012ApJ...747L..17C,2021ApJ...914...64T}. In this Section's discussion, we assume that the CSM velocity and the FS velocity is constant for simplicity.\par
            The following estimate adopts the unabsorbed X-ray luminosity as a proxy of the ionization source, and it is thus the overestimate of the ionization status by omitting the effect of the X-ray attenuation. It is thus appropriate for our main aim, i.e., to check the possible effect of the ionization to see if it would alter the conclusions obtained by the models assuming the effectively neutral CSM (Section \ref{sec:Application}).\par
            The ionization parameter is expressed as follows:
            \begin{equation}\label{eq:xi_general}
                    \xi\ [\mathrm{erg\ cm\ s^{-1}}]=\frac{L_{\mathrm{X,soft}}}{n_{\mathrm{e}} r^{2}},
            \end{equation}
            where $L_{\mathrm{X,soft}}$ and $n_{\mathrm{e}}$ are the unabsorbed soft X-ray luminosity and the number density of electrons in the unshocked CSM. The energy of photons that ionize the K-shell of C or O, as our main interest here, is mainly in the soft X-ray band.\par
            For C or O, the K-shell electrons are lost when $\xi\gtrsim100$ through the ionization by soft X-rays \citep{1969ApJ...156..943T,1976ApJ...206..847H,1982ApJS...50..263K,2012ApJ...747L..17C,2021ApJ...914...64T}. In the early epoch soon after the explosion, the soft X-ray luminosity is very large, so $\xi$ could become larger than 100; therefore C and O may be fully ionized. As time goes by, the X-ray luminosity decreases and $\xi$ will become smaller than 100 (see eq. \ref{eq:xi} and $L_{\mathrm{X,soft}}$ in captions of Figs. \ref{fig:photoionization_range_D'=0.1_D'=6.4}-\ref{fig:photoiongraphSN2022ablq}); the K-shell electrons of C or O will then be recombined, once the recombination timescale is elapsed (see eq. \ref{eq:recombinationtimescale}).\par
            Assuming $\rho\propto r^{-3}$ for the CSM density profile, the ionization parameter (eq. \ref{eq:xi_general}) is expressed as follows:
            \begin{equation}\label{eq:xi}
                \begin{split}
                    \xi
                    &=\frac{L_{\mathrm{X,soft}}}{n_{\mathrm{e}} r^{2}}\\
                    &\simeq \frac{L_{\mathrm{X,soft}}\ [\mathrm{erg\ s^{-1}}]}{\frac{ \rho_{\mathrm{CSM}\ [\mathrm{g\ cm^{-3}}]}}{2m_{\mathrm{p}}\ [\mathrm{g}]}r^{2}\ [\mathrm{cm^{2}}]}\\
                    &\simeq 19.2
                    \left(\frac{L_{\mathrm{X,soft}}}{10^{40}\ \mathrm{erg}}\right)
                    \left(\frac{r}{10^{15}\ \mathrm{cm}}\right)
                    \left(\frac{D'V_{\mathrm{sh}}}{V_{\mathrm{sh}}-V_{\mathrm{CSM}}}\right)^{-1}.
                \end{split}
            \end{equation}
             The recombination timescale $t_{\mathrm{rec},j}$ of the element $j$ is expressed as follows \citep[e.g.,][]{1982A&A...111..140F}:
            \begin{equation}\label{eq:recombinationtimescale}
                \begin{split}
                    t_{\mathrm{rec},j}\ [\mathrm{s}]
                    &=(n_{\mathrm{e}}\alpha(T_{\mathrm{e}}))^{-1}\\
                    &\simeq\left(\frac{\rho_{\mathrm{CSM}}}{2m_{\mathrm{p}}}\alpha(T_{\mathrm{e}})\right)^{-1}\\
                    &\simeq 4\times 10^{3} 
                    \left(\frac{D'}{1}\right)^{-1}
                    \left(\frac{r}{10^{15}\ \mathrm{cm}}\right)^{3}
                    \left(\frac{T_{\mathrm{e}}}{10^{6}\ \mathrm{K}}\right)^{0.5}
                    Z_{j}^{-2},
                \end{split}
            \end{equation}
            where $\alpha(T_{\mathrm{e}})$ is the recombination coefficient. We note that eq. \ref{eq:recombinationtimescale} ignores the CSM velocity, i.e., the recombination timescale is overestimated. However, it is justified for our purpose of confirming the assumption of the effectively neutral CSM. According to eq. \ref{eq:xi}, when $L_{\mathrm{X,soft}}$ is fixed (i.e., if the time since the explosion is fixed), $\xi$ is smaller for the inner region. In addition, according to eq. \ref{eq:recombinationtimescale}, the recombination timescale $t_{\mathrm{rec},j}$ is shorter for the inner region. Furthermore, the optical depth of photoelectric absorption above the radius $r$ is expressed as $\tau\propto r^{-2}$. Summarising these conditions, the ionization state at the FS front essentially determines whether photoelectric absorption is important or not (see Figure \ref{Photoionization_Changing_Overview}). Therefore, in the following discussion, we focus on the ionization state at the FS front.\par
            Here, we discuss the ionization state at radius $r$. We define $t_{\mathrm{sh,arr}}$ as the time for the FS to reach to $r$; 
            \begin{equation}\label{eq:time_sh_arr}
                t_{\mathrm{sh,arr}}=\frac{r}{V_{\mathrm{sh}}}.
            \end{equation}
            At time $t_{\mathrm{sh,arr}}$, the criterion that photoelectric absorption is important is expressed as follows:
            \begin{equation}\label{eq:criterion_recombination}
                t_{\mathrm{rec},j}(r) + t(\xi\simeq100;\ \mathrm{at}\ r) < t_{\mathrm{sh,arr}},
            \end{equation}
            where $t(\xi\simeq100;\ \mathrm{at}\ r)$ is the time since the explosion when $\xi$ at $r$ has decreased to 100. If this relation is satisfied, the CSM at $r$ is recombined before the FS reaches there, providing the source of photoelectric absorption. 
            We note that the recombination time is negligible as compared to $t(\xi\simeq100;\ \mathrm{at}\ r)$ in the early phase, 
            and thus for our purpose the recombination timescale can be practically omitted from eq. 
            \ref{eq:criterion_recombination}. \
            We note that the critical value set as $\xi\simeq100$ is a rough estimate, and indeed can be an underestimate \citep{1976ApJ...206..847H}; we thus adopt a range of $\xi=100-200$ in the subsequent discussion.\par
            In the earliest epoch of SNe Ibn/Icn, it is possible that eq. \ref{eq:criterion_recombination} is not satisfied because the soft X-ray luminosity could be very large. Then, the CSM is likely fully ionized at the FS front, and whether the soft X-rays can escape or is not determined by photoelectric absorption but by Compton scattering. This will generally lead to more luminous soft X-ray emission than computed in Sections \ref{sec:Results} and \ref{sec:Application} in the earliest epoch.\par
            If eq. \ref{eq:criterion_recombination} is satisfied at the FS front before the soft X-ray peak time computed without the ionization effect (see Sections \ref{sec:Results} and \ref{sec:Application}), the `real' soft X-ray LC, taking into account the ionization effect, could show double peaks as demonstrated in Figure \ref{DoublePeakLC_Overview}. In this case, the soft X-rays can escape initially through the fully-ionized CSM in the earliest phase. As time passes, the unabsorbed X-ray emission becomes weak and the ions in the unshocked CSM start recombining. Then, the optical depth suddenly increases due to the photoelectric absorption, leading to a sharp dimming of the soft X-ray LC. Once the FS front becomes effectively neutral, the optical depth to the photoelectric absorption will simply decrease, thus the soft X-ray LC will rise again toward the (second) peak.\par
            
        \begin{figure}[t]  
            \centering
            \includegraphics[keepaspectratio, scale=0.4]{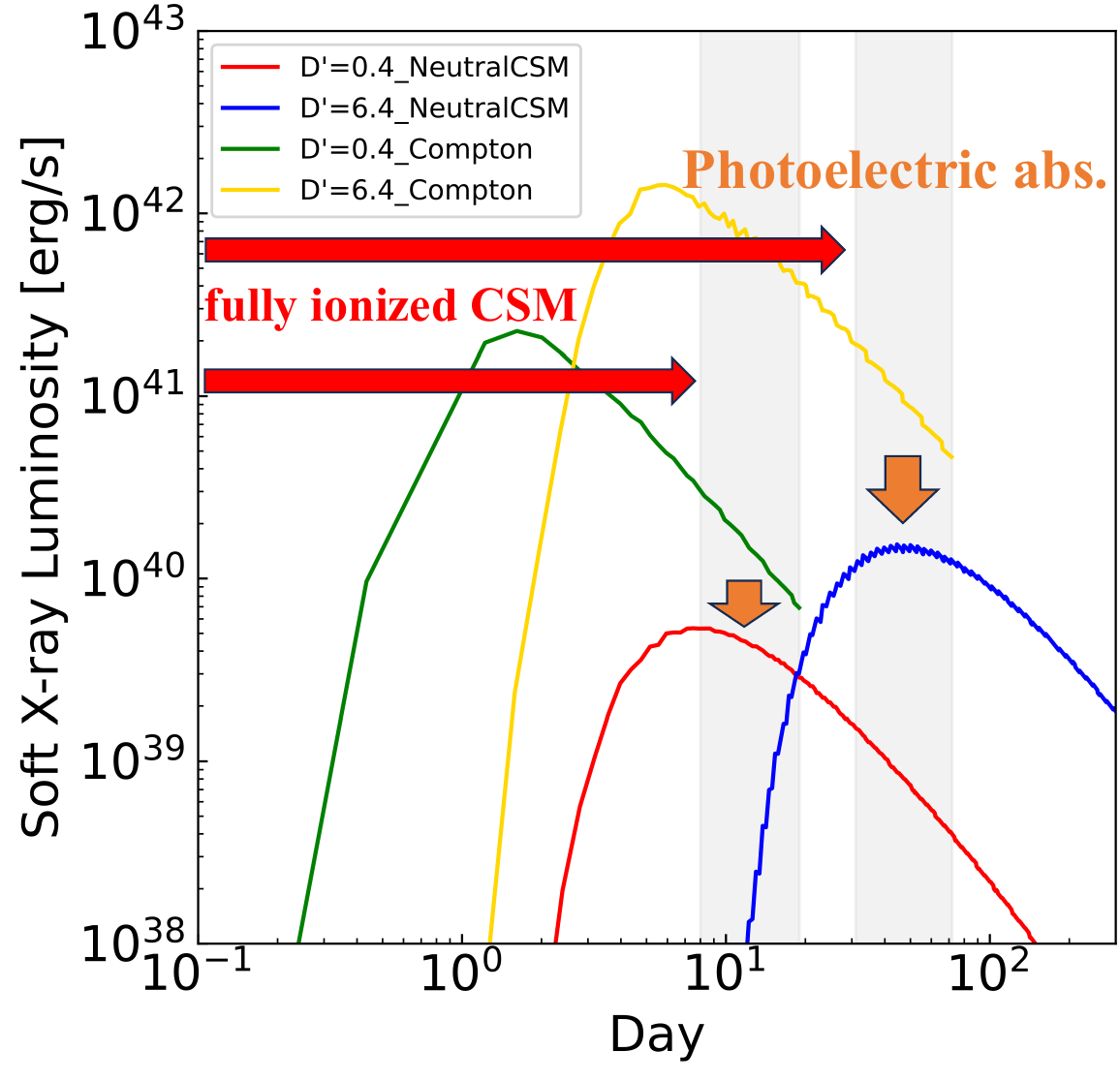}
            \caption{The soft X-ray LCs the models with for $D'=0.4$ (red and green lines) and $6.4$ (yellow and blue lines), adopting X (He)=0.5 for CSM composition. The green and yellow lines show the synthetic LCs taking into account only Compton scattering. The red and blue lines are the synthetic LCs assuming the effectively neutral CSM. The grey area corresponds to the `recombination' period, as estimated with $\xi=100-200$ at the unshocked CSM just ahead of the FS.\label{softx-ray_for_photoion_with_compton}}
        \end{figure}

        \begin{figure*}[t]
            \begin{minipage}[b]{0.1\linewidth}
            \centering
            \includegraphics[keepaspectratio, scale=0.4]{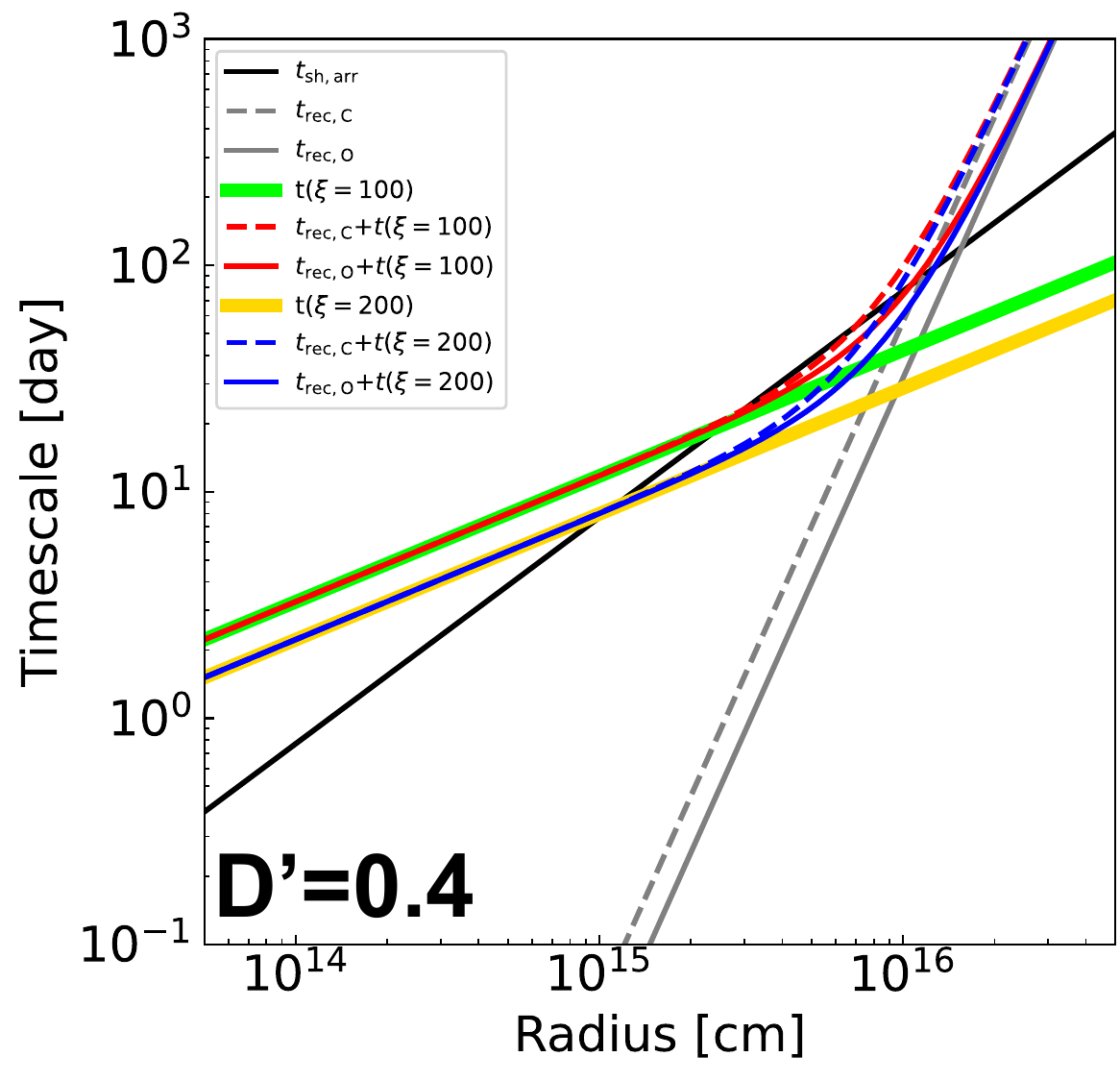}
            \end{minipage}
            \hfill
            \begin{minipage}[b]{0.5\linewidth}
            \centering
            \includegraphics[keepaspectratio, scale=0.4]{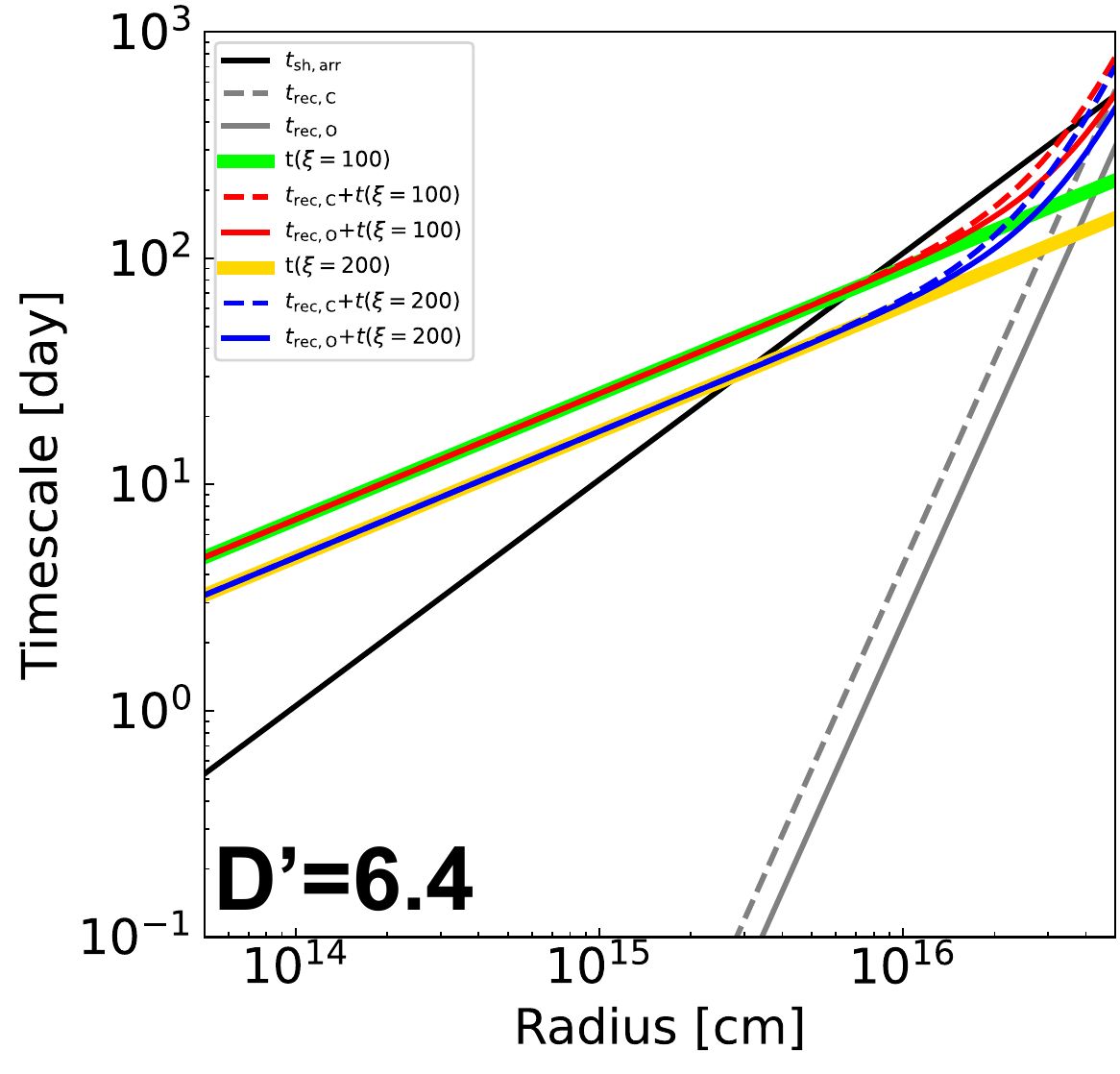}
            \end{minipage}
                \caption{
                The diagnosing plots for the ionization state in unshocked CSM, adopting $D'=0.4$ (left) and $6.4$ (right). The lines correspond to various timescales (see the legend). 
                We assume $T_{\mathrm{e}}=10^{6}\ \mathrm{K}$ \citep[e.g.,][]{1996ApJ...461..993F}.
                For $D'=0.4$ (left), we adopt the following conditions; $V_{\mathrm{sh}}=1.5\times 10^{9}\ \mathrm{cm\ s^{-1}}$, and $L_{\mathrm{X,soft}}=2.0\times10^{40}(\frac{t}{10\ \mathrm{days}})^{-1.8}\ \mathrm{erg\ s^{-1}}$ (see Fig. \ref{softx-ray_for_photoion_with_compton}). For $D'=6.4$ (right), we adopt the following conditions; $V_{\mathrm{sh}}=1.1\times 10^{9}\ \mathrm{cm\ s^{-1}}$, and $L_{\mathrm{X,soft}}=2.0\times10^{40}(\frac{t}{100\ \mathrm{days}})^{-1.8}\ \mathrm{erg\ s^{-1}}$ (see Fig. \ref{softx-ray_for_photoion_with_compton}).
                \label{fig:photoionization_range_D'=0.1_D'=6.4}}
        \end{figure*}

        \begin{figure*}[t]
            \begin{minipage}[b]{0.45\linewidth}
            \centering
            \includegraphics[keepaspectratio, scale=0.40]{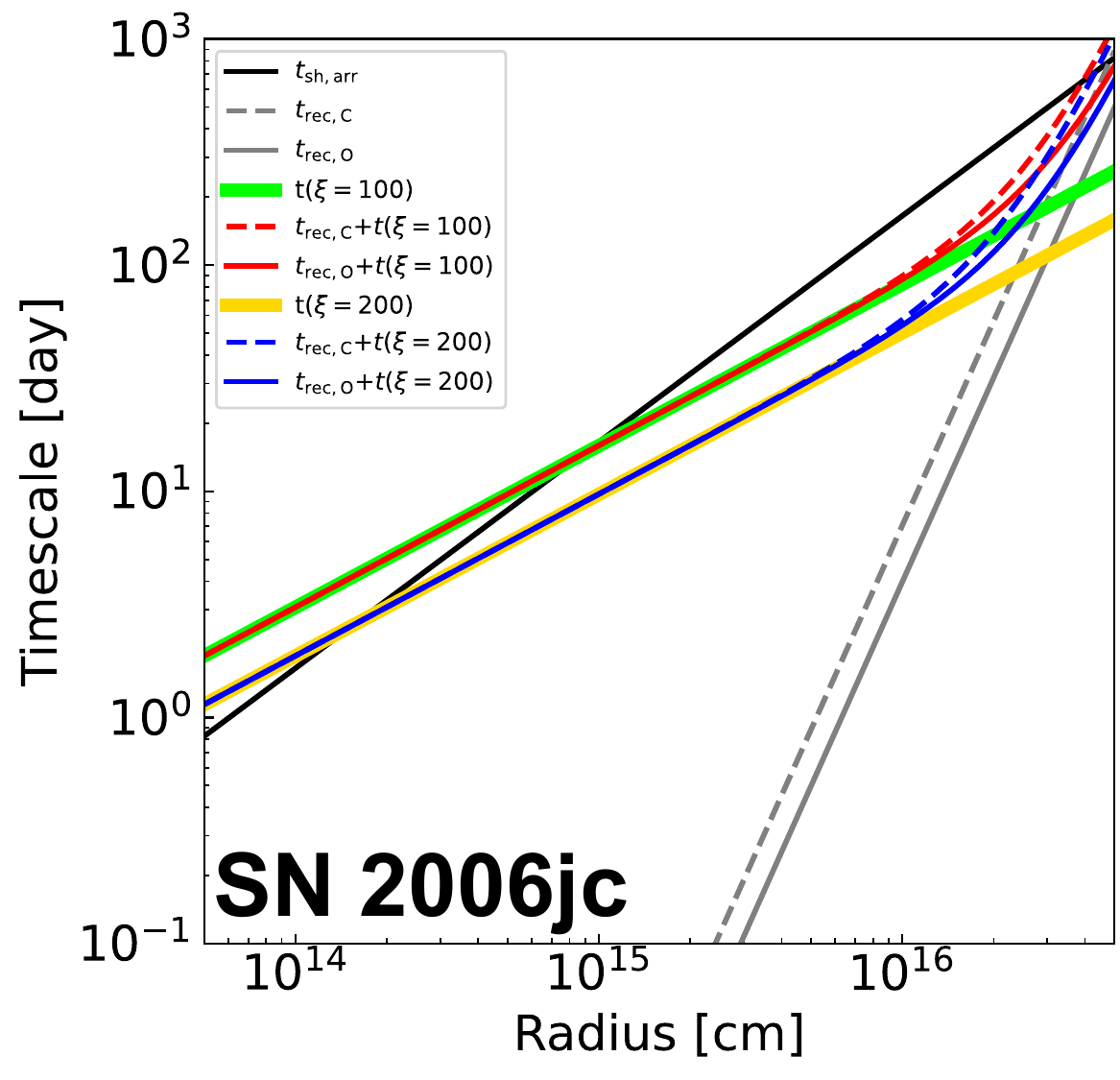}
            \end{minipage}
            \hfill
            \begin{minipage}[b]{0.45\linewidth}
            \centering
            \includegraphics[keepaspectratio, scale=0.45]{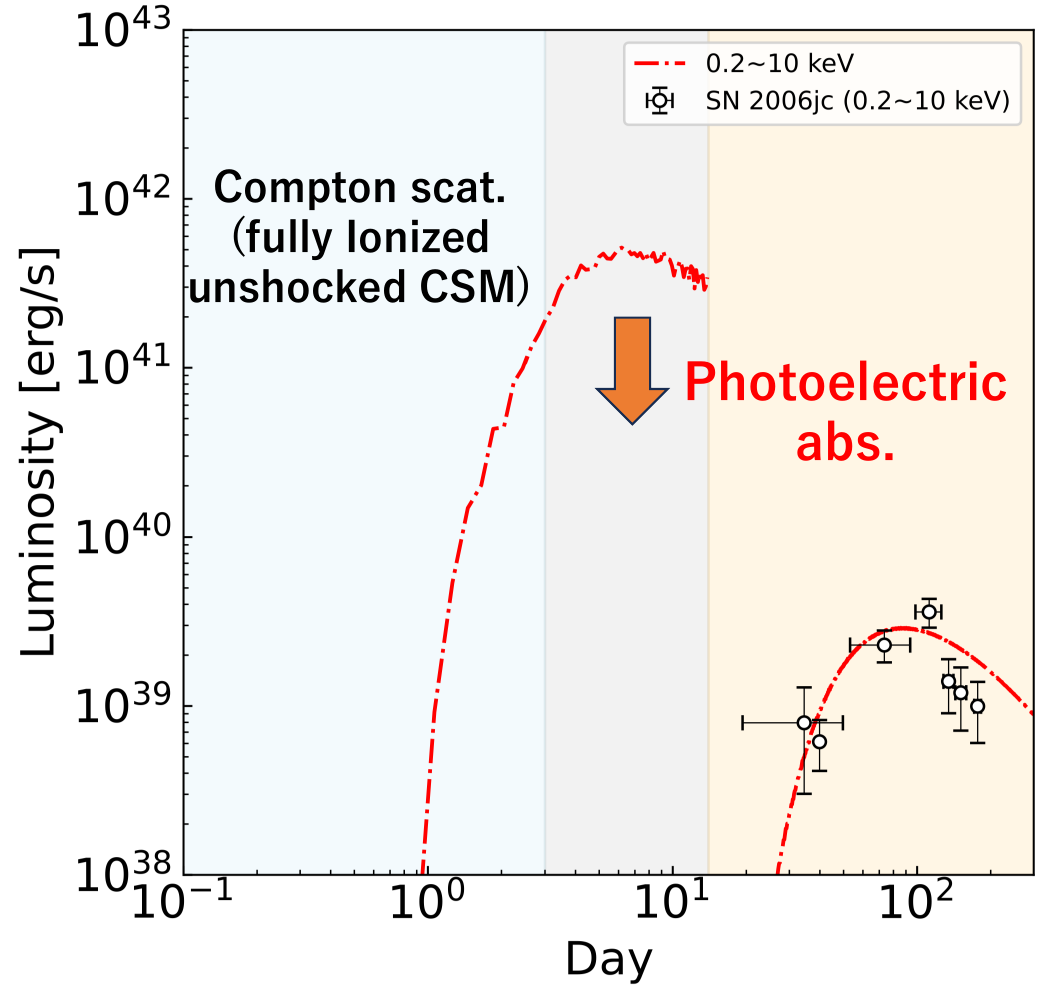}
            \end{minipage}
                \caption{
                The left panel shows the diagnosing plot for the ionization state in the unshocked CSM of SN 2006jc (see the caption of Fig. \ref{fig:photoionization_range_D'=0.1_D'=6.4}). We adopt the following conditions; $V_{\mathrm{sh}}=0.7\times 10^{9}\ \mathrm{cm\ s^{-1}}$, and $L_{\mathrm{X,soft}}=2.0\times10^{40}(\frac{t}{100\ \mathrm{days}})^{-1.4}\ \mathrm{erg\ s^{-1}}$. The right panels shows the model for the absorbed (observed) soft X-ray LC of SN 2006jc (red lines), as compared to the observed data points for SN 2006jc taken from \citet{2008ApJ...674L..85I}. In the earliest phase (blue area), the soft X-ray luminosity is computed by taking into account only Compton scattering. In the orange area, the soft X-ray luminosity is computed by taking into account photoelectric absorption on the effectively neutral CSM. The grey area corresponds to the recombination period as estimated with $\xi=100-200$ for the critical ionization parameter.
                \label{fig:photoiongraphSN2006jc}}
        \end{figure*}

        \begin{figure*}[p]
            \begin{minipage}[b]{0.45\linewidth}
            \centering
            \includegraphics[keepaspectratio, scale=0.40]{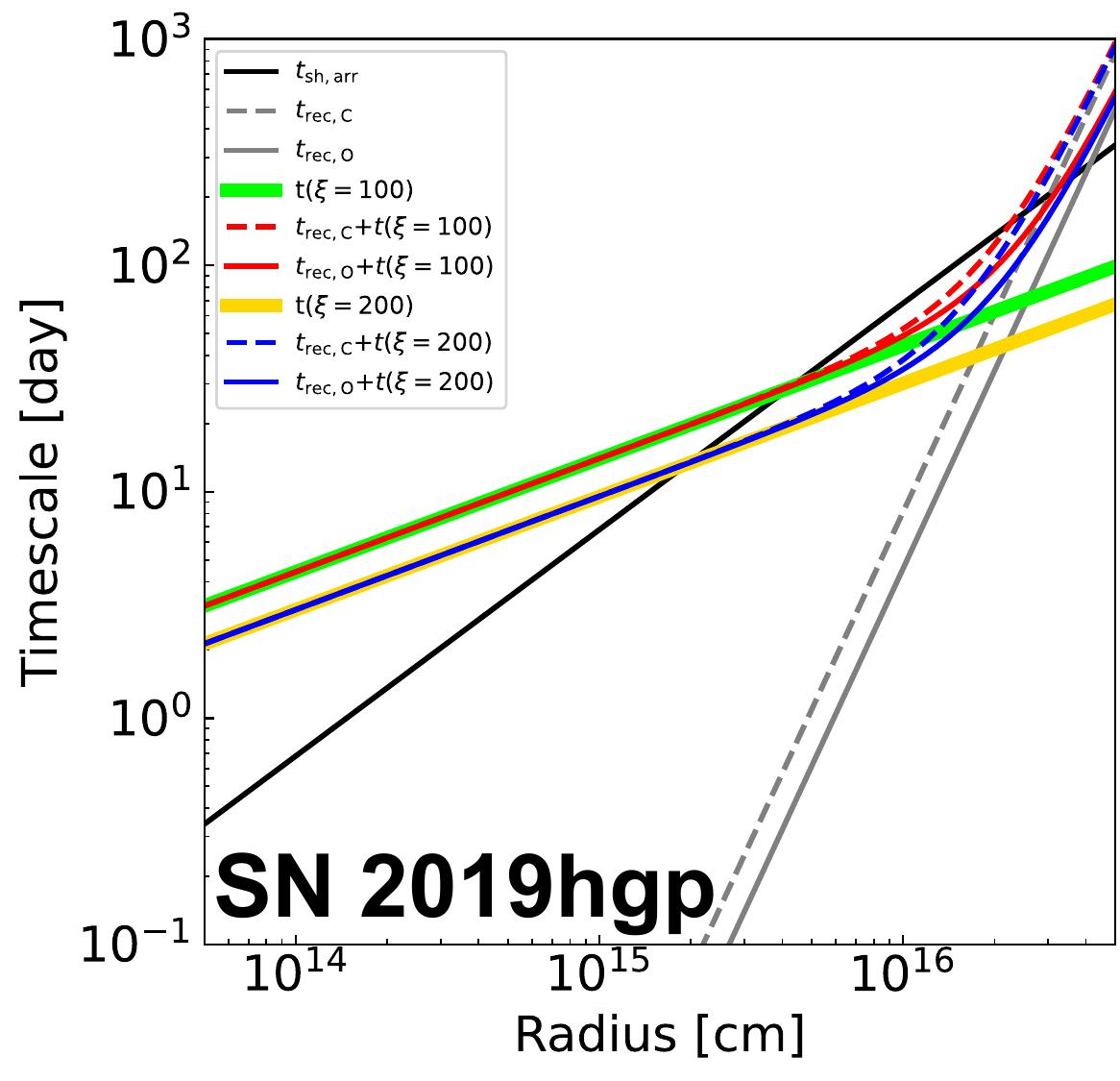}
            \end{minipage}
            \hfill
            \begin{minipage}[b]{0.45\linewidth}
            \centering
            \includegraphics[keepaspectratio, scale=0.45]{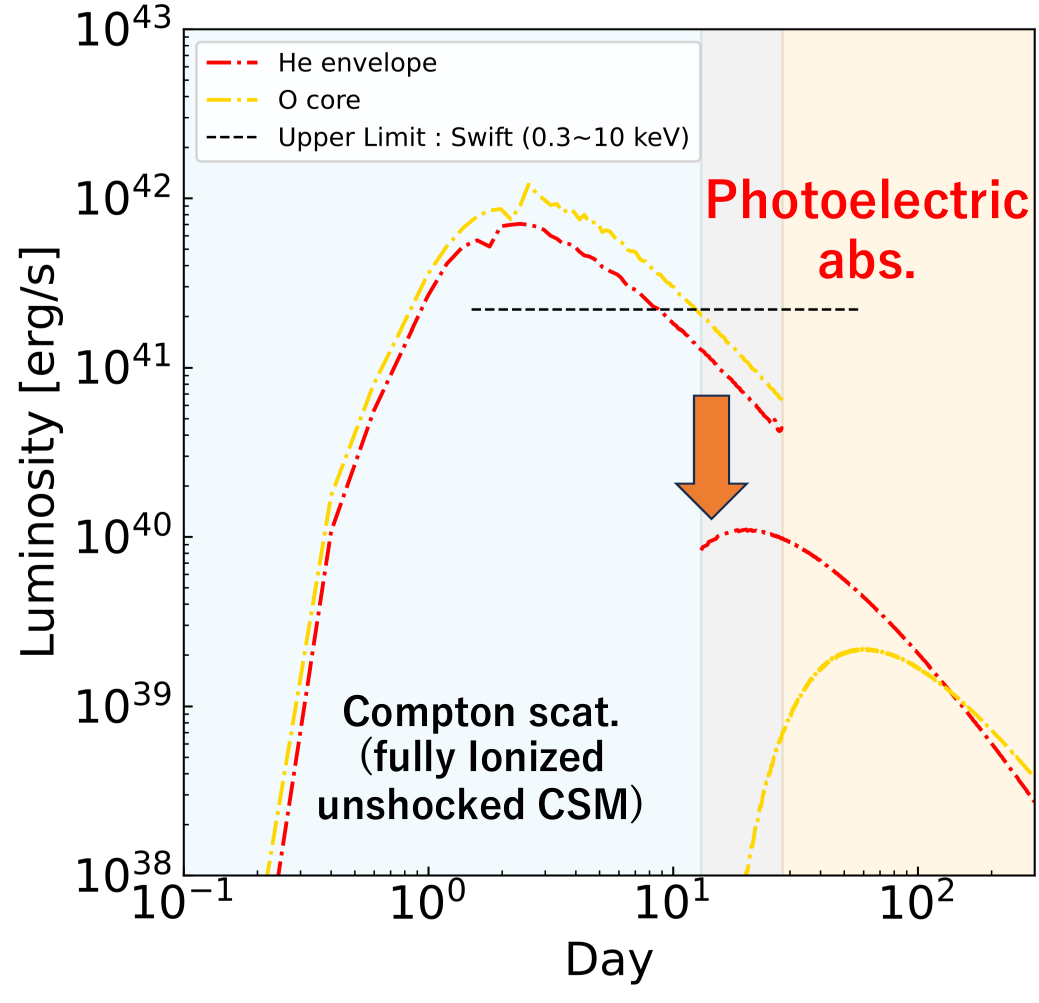}
            \end{minipage}
                \caption{
                The left panel shows the diagnosing plot for the ionization state in the unshocked CSM of SN 2019hgp. We adopt the following conditions; 
                $V_{\mathrm{sh}}=1.7\times 10^{9}\ \mathrm{cm\ s^{-1}}$, and $L_{\mathrm{X,soft}}=3.0\times10^{39}(\frac{t}{100\ \mathrm{days}})^{-1.7}\ \mathrm{erg\ s^{-1}}$.
                The right panels shows the model for the absorbed (observed) soft X-ray LC of SN 2019hgp adopting the `He-envelope' compositions (dashed-red) and the `O-core' compositions (dashed-yellow) (see Section \ref{subsubsec:SN2019hgp}). The upper limit of the soft X-ray luminosity of SN 2019hgp is taken from \citet{2022Natur.601..201G}. See the caption of Fig. \ref{fig:photoiongraphSN2006jc}. 
                \label{fig:photoiongraphSN2019hgp}}
        \end{figure*}

        \begin{figure*}[p]
            \begin{minipage}[b]{0.45\linewidth}
            \centering
            \includegraphics[keepaspectratio, scale=0.40]{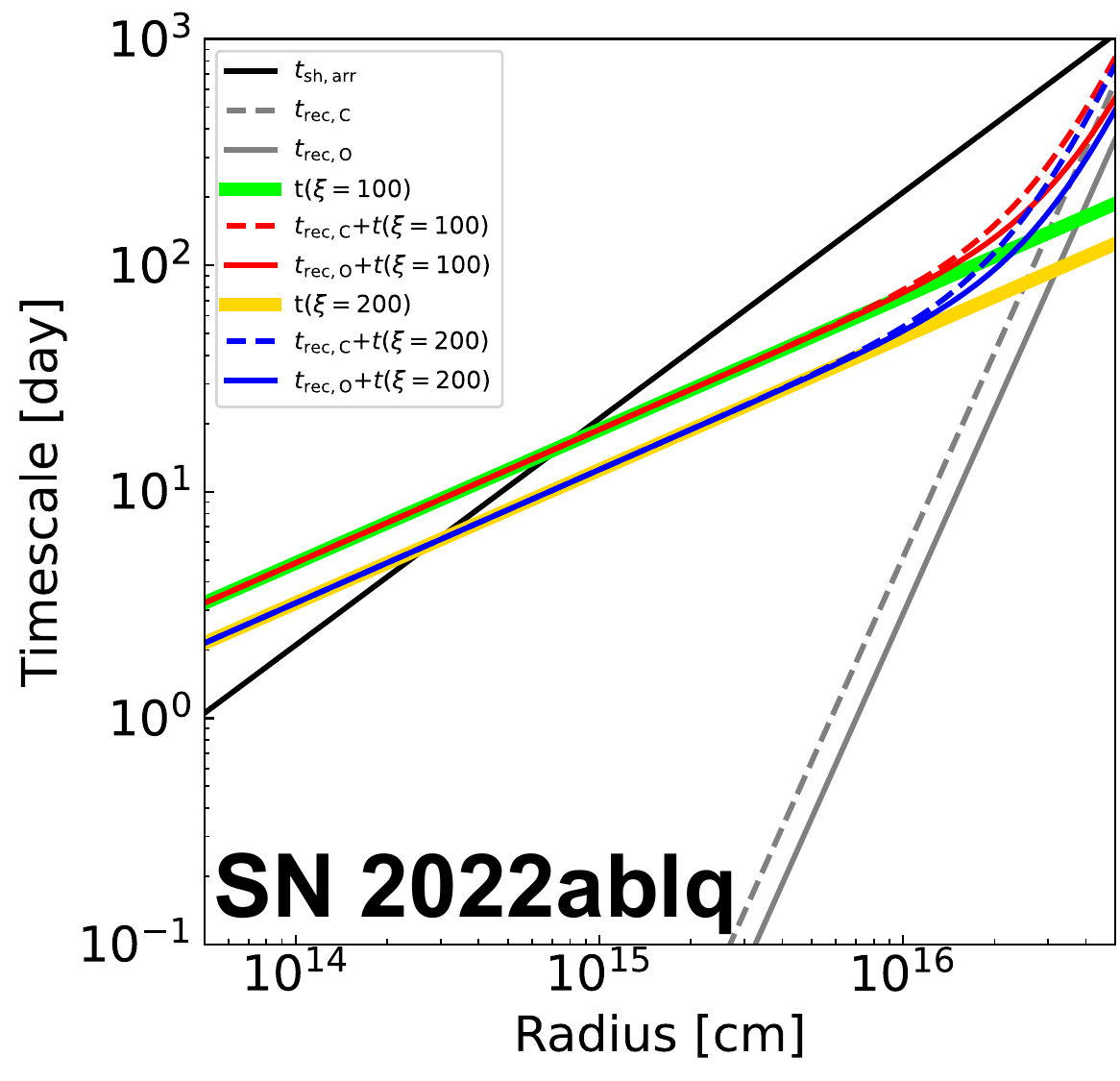}
            \end{minipage}
            \hfill
            \begin{minipage}[b]{0.45\linewidth}
            \centering
            \includegraphics[keepaspectratio, scale=0.45]{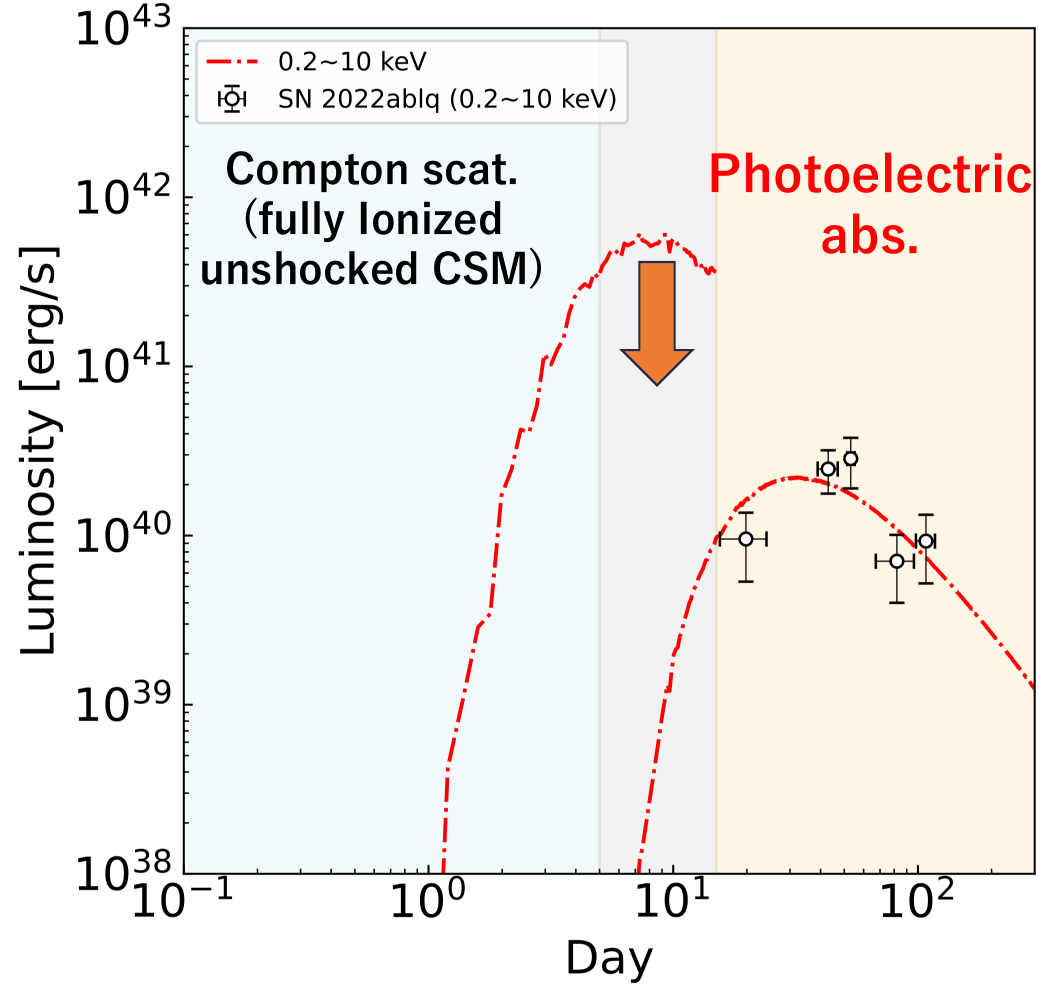}
            \end{minipage}
                \caption{
                The left panel shows the diagnosing plot for the ionization state in the unshocked CSM of SN 2022ablq. We adopt the following conditions; $V_{\mathrm{sh}}=0.55\times 10^{9}\ \mathrm{cm\ s^{-1}}$, and $L_{\mathrm{X,soft}}=1.2\times10^{40}(\frac{t}{100\ \mathrm{days}})^{-1.7}\ \mathrm{erg\ s^{-1}}$.
                The right panels shows the model for the absorbed (observed) soft X-ray LC of SN 2022ablq (red lines), as compared to the observed data points taken from \citet{2024arXiv240718291P}. See the caption of Fig. \ref{fig:photoiongraphSN2006jc}. 
                \label{fig:photoiongraphSN2022ablq}}
        \end{figure*}

        \subsection{Double-peaked soft X-ray light-curves}\label{subsec:photoiongeneral}
            We apply the photoionization model in Section \ref{subsec:photoionization} to the models with $D'=0.4$ and $6.4$ (Fig. \ref{softx-ray_for_photoion_with_compton} for the `neutral' CSM). The timescales mentioned in Section \ref{subsec:photoionization} are shown in Figure \ref{fig:photoionization_range_D'=0.1_D'=6.4}. Figure \ref{fig:photoionization_range_D'=0.1_D'=6.4} shows that eq. \ref{eq:criterion_recombination} is satisfied in the time window at $>8$ ($\xi=200$) or $>19$ ($\xi=100$) days ($D'=0.4$), or $>31$ ($\xi=200$) or $>72$ ($\xi=100$) days ($D'=6.4$).
            The peak dates computed for the `neutral' CSM are in these time windows (Fig. \ref{softx-ray_for_photoion_with_compton}; $\sim8$ days for $D'=0.4$ and $\sim50$ days for $D'=6.4$). The analysis here demonstrates that the assumption of the neutral CSM is not necessarily correct in the early phase, and the effect of the ionization should be checked in studying individual objects. Further, the ionization effect can create a unique soft X-ray evolution, i.e., the double peaks (or a peak followed by a shoulder), demonstrating the importance of prompt soft X-ray observations in the early phase (see Fig. \ref{softx-ray_for_photoion_with_compton}).\par

        \subsection{Revisiting the soft X-ray emissions from SNe 2006jc, 2019hgp, and 2022ablq}\label{subsec:doublepeakofSN2006jcSN2019hgp}
            The soft X-ray data of SN 2006jc in the range of $\sim$20-200 days were fit in Section \ref{subsec:2006jc}, assuming that the unshocked CSM was already recombined in these epochs. In estimating the effect of the ionization for SN 2006jc, we approximate its unabsorbed soft X-ray luminosity as $L_{\mathrm{X,soft}}=2\times10^{40}(\frac{t}{100\ \mathrm{days}})^{-1.4}\ \mathrm{erg\ s^{-1}}$ (Fig. \ref{SN2006jc,xray}) and its shock velocity as $V_{\mathrm{sh}}=0.7\times 10^{9}\ \mathrm{cm\ s^{-1}}$ (as given by SNEC).
            Adopting the same model parameters with those shown in Figure \ref{SN2006jc,xray}, the diagnosing plots for the ionization status are shown in right panel in Fig. \ref{fig:photoiongraphSN2006jc}.
            We estimate the the inner-shell electrons of C and O in the FS front are recombined between the first 3 ($\xi=200$) and 14 ($\xi=100$) days since the explosion, after which photoelectric absorption is expected to fully operate as was assumed in Section \ref{subsec:2006jc}. In addition to this confirmation, the analysis here provides an interesting implication; our model predicts that a bright soft X-ray might have been associated with SN 2006jc within the first 3 ($\xi=200$) or 14 ($\xi=100$) days (left panel in Fig. \ref{fig:photoiongraphSN2006jc}), if it were observed in the soft X-ray band promptly just after the explosion.\par
            We have done the same exercises for SNe 2019hgp and 2022ablq (Figs. \ref{SNeIcn,x-ray} and \ref{SN2022ablq,xray}). We adopt $L_{\mathrm{X,soft}}=3.0\times10^{39}(\frac{t}{100\ \mathrm{days}})^{-1.7}\ \mathrm{erg\ s^{-1}}$ and $V_{\mathrm{sh}}=1.7\times 10^{9}\ \mathrm{cm\ s^{-1}}$ for SN 2019hgp; $L_{\mathrm{X,soft}}=1.2\times10^{40}(\frac{t}{100\ \mathrm{days}})^{-1.7}\ \mathrm{erg\ s^{-1}}$ and $V_{\mathrm{sh}}=0.55\times 10^{9}\ \mathrm{cm\ s^{-1}}$ for SN 2022ablq. The relatively low FS velocity adopted here is consistent with the shock velocity of SN 2022ablq estimated by \citet{2024arXiv240718291P}. We find that photoelectric absorption starts attenuating the soft X-ray at $\sim5$ ($\xi=200$) or $\sim 15$ ($\xi=100$) days for SN 2022ablq (Fig. \ref{fig:photoiongraphSN2022ablq}. The phase is before the first X-ray detection, confirming the consistency of our derived parameters for SN 2022ablq in Section \ref{subsec:2022ablq} with the X-ray observation.)\par

            A possible issue is raised for the ionization effect for SN 2019hgp. We find that photoelectric absorption starts attenuating the soft X-ray between $13$ and $28$ days for SN 2019hgp (see Fig. \ref{fig:photoiongraphSN2019hgp}), thus the expected early `bright' phase overlaps with part of the periods when it was observed by Swift resulting in non-detection \citep{2022Natur.601..201G}. With the ionization effect considered here, we indeed expect that it should have been detected (Fig. \ref{fig:photoiongraphSN2019hgp}). This might demonstrate several limitations in the present analysis, including the followings; there are uncertainty in the physical parameters estimated by the optical LC modeling. The earliest epoch at $\lsim 4$ days might be contaminated by an LC component other than the interaction, and it is beyond the applicability of the optical LC modeling presented in \citet{2023A&A...673A..27N} (see Fig. 4 of \citealt{2023A&A...673A..27N}); without detection, deriving the X-ray luminosity upper limit involves uncertainties (e.g., assumed SED); we might overestimate the ionization effect by ignoring the effect of the attenuation of X-rays. To settle this issue, more detailed calculations including the attenuation effect are needed, together with detection of the hard X-rays (to accurately derive the physical parameters such the the CSM density; see Section \ref{subsec:obspossivility}) and the soft X-rays (to calibrate the model).

\section{Discussion} \label{sec:Discussion}
    \subsection{The importance of hard X-ray observation}\label{subsec:obspossivility}
        Deriving the CSM composition (e.g., SNe 2006jc and 2022ablq in Section \ref{subsec:2006jc}) is influenced by the assumed CSM density distribution. This is because the rising properties of the soft X-ray LC are controlled by the column density of heavier elements that is determined by the CSM density and composition as shown in eq. \ref{obsxray_2}. Here is where the hard X-ray LC provides an independent and powerful diagnostics;  even in the rising phase of the soft X-ray LC, the hard X-rays can already be in the optically-thin phase, and it can provide a robust measure of the CSM density parameters ($D'$ and $s$). The CSM abundance can be then derived through the soft-X-ray LC analysis. Given its nature as a bolometer of the SN-CSM interaction energetics, the hard X-ray LC is a more direct trace of the CSM density properties than in the optical LC analysis. \par
        As shown in Fig. \ref{SNeIcn,x-ray}, NuSTAR provides a sensitivity that is expected to be sufficient to detect nearby SNe Ibn/Icn in hard X-rays, if the observation is performed promptly after the explosion. This will provide key contribution in studying the progenitor of SNe Ibn/Icn and their final activities.\par

    \subsection{Possible effects of inverse Compton scattering}\label{subsec:inversecompscat_opt}
        According to the previous study \citep{2022ApJ...927...25M}, the main energy source of the optical photon is provided by the inward-going X-rays originally emitted at the FS, which are then reprocessed into the optical photons either in the FS, RS or the unshocked ejecta. The outward-going optical photons then inversely pass through the outer layers such as the RS or FS, and may experience inverse Compton scattering by these regions and be energized to X-rays. However, when a non-negligible fraction of X-rays can escape the shocked regions, the system should also be optically thin to optical photons. Therefore, the contribution of the X-ray photons created through inverse Compton scattering by thermal electrons will not be a main contributor to the X-ray LC.\par
        In addition, we note that our model dose not treat non-thermal (relativistic) electrons which are generated at the FS front thorough the DSA (Diffusive shock acceleration) mechanism. In some situations, inverse Compton scattering of optical photons by these non-thermal electrons could contribute to the X-ray emission \citep[see e.g.,][about charged particle acceleration at the shock front]{1978MNRAS.182..147B,1978MNRAS.182..443B,1983RPPh...46..973D,1998ApJ...509..861F,2003LNP...598..171C,2017hsn..book..875C,2012ApJ...758...81M,2013ApJ...762...14M,2019ApJ...885...41M,2019ApJ...874...80M}. In an extreme circumstance, synchrotron emission could directly contribute to X-ray emission \citep[e.g.,][]{2019ApJ...874...80M}. While the particle acceleration process \citep[e.g.,][]{1949PhRv...75.1169F,1978MNRAS.182..147B,1978MNRAS.182..443B,1983RPPh...46..973D} and the efficiency are still under debate, for the high-density CSM considered here, this will be overwhelmed by the thermal X-ray emission \citep[e.g.,][]{2014ApJ...785...95M}. One possible exception is the late-time tail phase, where the efficiency of the thermal X-ray emission decreases due to the rapidly decreasing CSM density for SNe Ibn/Icn; this is an issue we plan to investigate in the future.

\section{Summary}\label{sec:summary}
    In the present study, we have developed a X-ray LC model for SNe Ibn/Icn, taking into account the rapid cooling behind the FS as a post-process (Section \ref{sec:method}). Through studying the parameter dependence on the resulting emission (Section \ref{sec:Results}), our findings can be summarized as follows: 
\begin{enumerate}
\item  The hard X-ray LC provides a robust measure of the explosion properties ($E_\mathrm{ej},\ M_\mathrm{ej}$) and the CSM density distribution ($\rho_{\mathrm{CSM}}$), as a tracer of the SN-CSM interaction energetics.
\item The soft X-ray LC is important for revealing the CSM composition, thanks to its rising LC properties controlled by the photoelectric absorption in the unshocked CSM.
\item Therefore, the broad-band X-ray observation is required for robustly revealing the nature of the progenitor and its final activities.
\end{enumerate}
\par
    We have then applied our X-ray LC model to the observational data of SNe 2006jc, 2019hgp, and 2022ablq. Under the assumption of the effectively neutral CSM (where the inner-shell electrons of carbon/oxygen are bound), the results can be summarized as follows: 
\begin{enumerate}
\item  The X-ray LC of SN 2006jc can be explained by the physical parameters derived from the optical LC modeling; $M_\mathrm{ej}=6\ \mathrm{M_\odot}$, $E_\mathrm{kin,ej}=0.8\ \mathrm{B}$, $V_\mathrm{CSM}=3000\ \mathrm{km\ s^{-1}}$, $s=3$ and $D'=4.0$. In addition, the X-ray LC of SN 2022ablq can be explained by the physical parameters of typical SNe Ibn derived from the optical LC modeling; $M_\mathrm{ej}=6\ \mathrm{M_\odot}$, $E_\mathrm{kin,ej}=0.6\ \mathrm{B}$, $V_\mathrm{CSM}=1500\ \mathrm{km\ s^{-1}}$, $s=3$ and $D'=4.0$. These demonstrate a general consistency between the optical and X-ray properties on the basis of the SN-CSM interaction model, supporting the interaction scenario to explain the nature of SNe Ibn (and Icn).
\item Our model indicates difference in the CSM compositions between SNe 2006jc and 2022ablq despite their both being SNe Ibn; (He, C, O)=(0.4, 0.3, 0.3) for SN 2006jc and (He, C, O)=(0.95, 0.025, 0.025) for 2022ablq, respectively (see Appendix \ref{subsec:comperror} about the robustness in deriving the CSM composition). This indicates that even within the same SN Ibn class, there could be a diversity in the degree of the mass loss and envelope stripping; the mass loss of the SN 2006jc progenitor might have reached down nearly to the bottom of the He envelope, while SN 2022ablq progenitor might have experienced a less extreme stripping where the stripping has been reached to the middle of the He layer. 
\item However, the above conclusion on the CSM composition should not be over-interpreted, and will require further confirmation; it depends on the assumed CSM density distribution. A more robust estimate of the CSM composition can be obtained by adding the hard X-ray observation and modeling, highlighting the importance of the prompt hard X-ray follow-up observation. 
\item The soft X-ray upper limit for SN 2019hgp as obtained by Swift is consistent with the interaction model, and indeed consistent with any CSM compositions examined in the present work.
\item While the hard X-ray observation has not been reported for SN 2019hgp, our model predicts that it must have been detected if a prompt observation just after the discovery were conducted by NuSTAR. 
\end{enumerate} 
\par
    The assumption that the inner-shell electrons of carbon/oxygen are bound affects the strength of photoelectric absorption. We have further explored the effect of photoionization of the unshocked CSM by the X-ray emitted from the interacting region on the emerging soft X-rays:
\begin{enumerate}
\item In a few days since the explosion, we may indeed expect strong soft X-ray emission; the unshocked CSM just ahead of the FS is expected to be fully ionized due to high X-ray luminosity. 
\item As combined with `the second peak', as created through suppression of the soft X-ray once the CSM is recombined and then the rising LC as the column density decreases, we show that SNe Ibn/Icn may show double-peaked soft X-ray LC. 
\item For the models adopted for SNe 2006jc and 2022ablq, we have confirmed that the assumption of the effectively neutral CSM is appropriate in the epochs where the soft X-ray data are available (i.e., in the second peak). 
\item In addition to the hard X-ray observation, the prompt soft X-ray observation within a few days can thus be a new window to study the nature of the SNe Ibn/Icn progenitors and their final activities. 
\item Indeed, our tentative analysis of SN 2019hgp, we estimate that soft X-ray should have been detected in the earliest phase despite the non-detection by Swift. This demonstrates a possibility of further refining the model and constraining the nature of SNe Ibn/Icn. For more accurate calculations of photoelectric absorption, the simple treatment of ionization states of ions in the present study should be updated, and the attenuation of X-rays in the radial direction, which is related to the photoionization of ions, needs to be included.
\end{enumerate}

\appendix
\section{Feedback of radiation loss on hydrodynamics}\label{subsec:radiationhydro}
    The present model does not consider feedback of radiative loss on hydrodynamics. To investigate the influence of radiative cooling on hydrodynamics, radiation hydrodynamic (RH) simulations using SNEC are conducted with the most extreme model with $D'=6.4$ (see Section \ref{sec:Results} for the other parameters). We note that the SNEC assumes complete thermal equilibrium and blackbody radiation which lead to an overestimation of the influence of the radiation loss. We find that the RH simulation by SNEC results in the 20\% reduction in the shock velocity and the 100\% enhancement in the gas density behind the shock, as compared to the adiabatic hydrodynamic simulation. The relation between X-ray luminosity and the properties behind the FS is expressed as follows:
    \begin{equation}\label{Xraypropto}
        L_{\mathrm{X}}\propto T^{0.5} \rho^{2} \mathrm{Vol}_{\mathrm{FS}}\propto V_{\mathrm{sh}}^{4}\rho^{2} \ ,
    \end{equation}
    where $T$, $\rho$, and $\mathrm{Vol}_{\mathrm{FS}}$ are gas temperature, mass density, and volume of the region shocked by the FS. According to eq. \ref{Xraypropto}, the resulting X-ray luminosity is increased by 60\% as compared to the adiabatic hydrodynamic simulation. Therefore, the synthetic unabsorbed X-ray luminosity in the present study might be underestimated by a factor of at most 1.6 by neglecting the feedback effect.\par
    The optical depth is expressed as follows:
    \begin{equation}\label{RH_feedback_opticaldepth}
        \begin{split}
        \tau_{\mathrm{pe}}, \tau_{\mathrm{comp}}
        &\propto r_{\mathrm{sh}}^{-s+1} \ ({\rm for} \ s>1)\\
        &\propto V_{\mathrm{sh}}^{-s+1} \ .
        \end{split}
    \end{equation}
    According to eq. \ref{RH_feedback_opticaldepth}, the RH simulation for the model with $D'=6.4$ and $s=3$ provides 60\% higher optical depth than that provided by the adiabatic hydrodynamics simulation. Given that the model examined here sets basically the upper limit on the feedback effect. our X-ray LC model in the main text (Section \ref{sec:method}) underestimates the optical depth by at most a factor of 1.6.

\section{Robustness in deriving the CSM composition}\label{subsec:comperror}
To discuss the robustness in deriving the CSM compositions in Sections \ref{subsec:2006jc} and \ref{subsec:2022ablq}, we show the synthetic soft X-ray LCs with several models with different CSM compositions, as compared to the data of SNe 2006jc and 2022ablq in Figs. \ref{SN2006jc,xray,variousabundance} and \ref{SN2022ablq,xray,variousabundance}, respectively. Thanks to the data point of SN 2006jc at $\sim$40 days since the explosion, we find that our X-ray LC model prefers the low helium composition models (e.g., 20 \%, 30 \%, and 40 \% models). In addition, considering the peak soft X-ray data point of SN 2006jc, we conclude that the 40 \% helium model is plausible for the CSM composition of SN 2006jc. Similar to the discussion of SN 2006jc, from the data points in the rising phase and the peak phase, we conclude that the 95 \% helium model is plausible for the CSM composition of SN 2022ablq.
        \begin{figure}[t]
        \centering
        \includegraphics[scale=0.6]{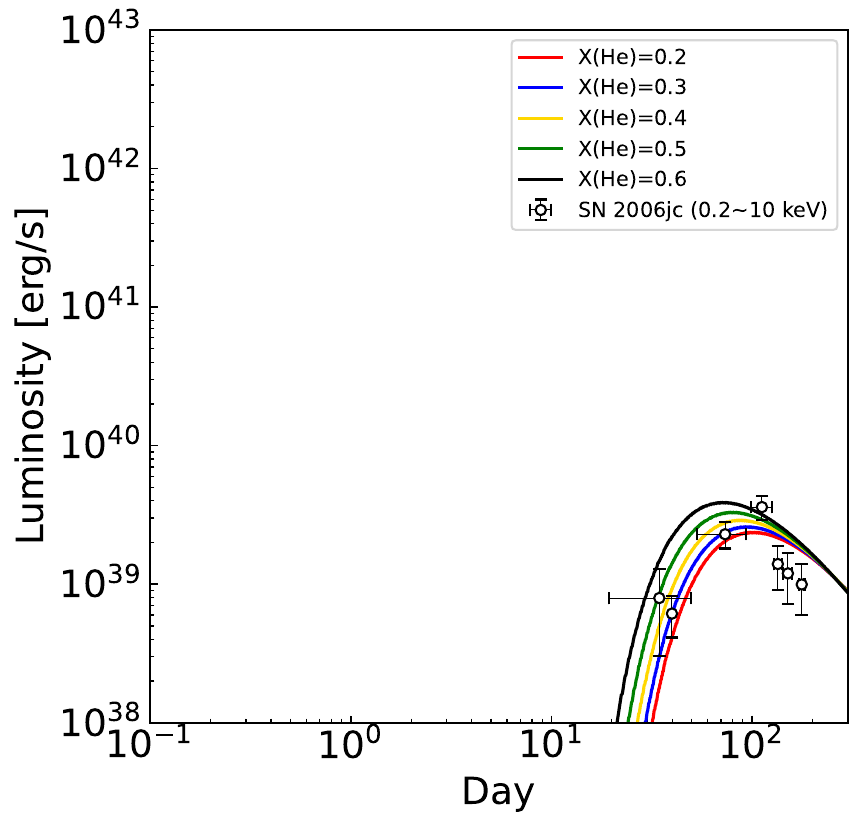}
        \caption{The synthetic soft (0.2-10 keV) X-ray LCs for SN 2006jc with the (effectively) neutral unshocked CSM, as compared to the X-ray data of SN 2006jc. Shown here are the models with (He, C, O)=(0.2, 0.4, 0.4) (red), (He, C, O)=(0.3, 0.35, 0.35) (blue), (He, C, O)=(0.4, 0.3, 0.3) (yellow), (He, C, O)=(0.5, 0.25, 0.25) (green), and (He, C, O)=(0.6, 0.2, 0.2) (black). The data points are from \citet{2008ApJ...674L..85I}.
        \label{SN2006jc,xray,variousabundance}}
        \end{figure}

        \begin{figure}[t]
        \centering
        \includegraphics[scale=0.6]{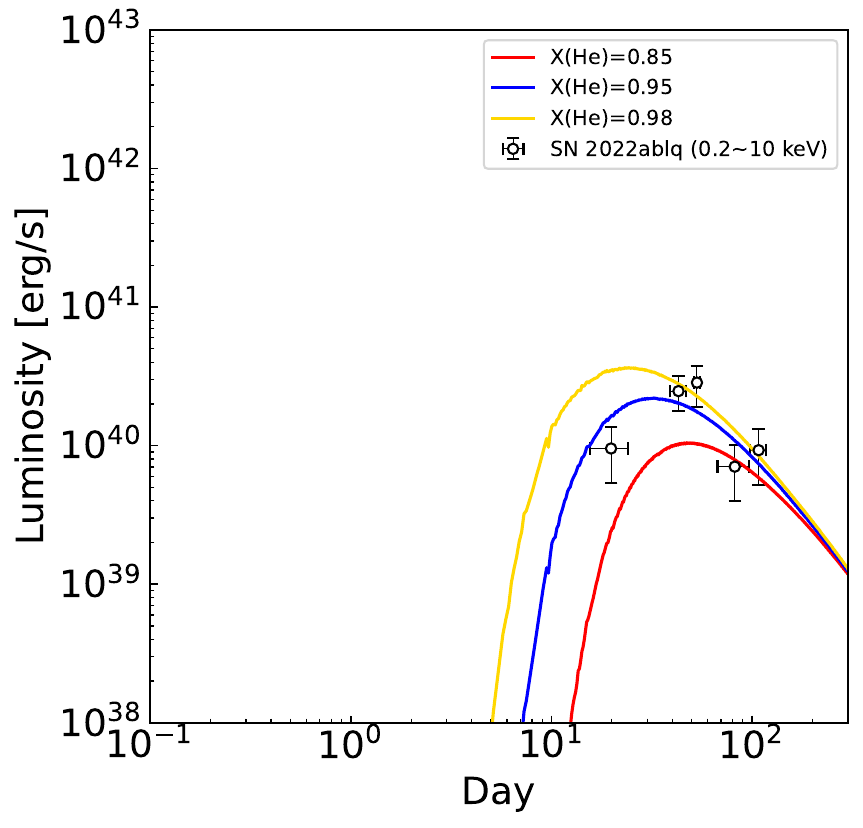}
        \caption{The synthetic soft (0.2-10 keV) X-ray LCs for SN 2022ablq with the (effectively) neutral unshocked CSM, as compared to the X-ray data of SN 2022ablq. Shown here are the models with (He, C, O)=(0.85, 0.075, 0.075) (red), (He, C, O)=(0.95, 0.025, 0.025) (blue), and (He, C, O)=(0.98, 0.01, 0.01) (yellow). The data points are from \citet{2024arXiv240718291P}. 
        \label{SN2022ablq,xray,variousabundance}}
        \end{figure}
\section*{acknowledgments}
The authors thank Shiu-Hang Lee, Tatsuya Matsumoto, Daichi Hiramatsu, Kenta Taguchi, Kohki Uno and Masaki Tsurumi for fruitful discussion. YI acknowledges supported by a grant from the Hayakawa Satio Fund awarded by the Astronomical Society of Japan.
KM acknowledges support from the JSPS KAKENHI grant JP20H00174, JP24H01810, and 24KK0070. 

\bibliography{manuscript}{}

\begin{thebibliography}{}
\expandafter\ifx\csname natexlab\endcsname\relax\def\natexlab#1{#1}\fi
\providecommand{\url}[1]{\href{#1}{#1}}
\providecommand{\dodoi}[1]{doi:~\href{http://doi.org/#1}{\nolinkurl{#1}}}
\providecommand{\doeprint}[1]{\href{http://ascl.net/#1}{\nolinkurl{http://ascl.net/#1}}}
\providecommand{\doarXiv}[1]{\href{https://arxiv.org/abs/#1}{\nolinkurl{https://arxiv.org/abs/#1}}}

\bibitem[{{Bell}(1978{\natexlab{a}})}]{1978MNRAS.182..147B}
{Bell}, A.~R. 1978{\natexlab{a}}, \mnras, 182, 147, \dodoi{10.1093/mnras/182.2.147}

\bibitem[{{Bell}(1978{\natexlab{b}})}]{1978MNRAS.182..443B}
---. 1978{\natexlab{b}}, \mnras, 182, 443, \dodoi{10.1093/mnras/182.3.443}

\bibitem[{{Brennan} {et~al.}(2024){Brennan}, {Sollerman}, {Irani}, {Schulze}, {Chen}, {Das}, {De}, {Fransson}, {Gal-Yam}, {Gkini}, {Hinds}, {Lunnan}, {Perley}, {Qin}, {Stein}, {Wise}, {Yan}, {Zimmerman}, {Anand}, {Bruch}, {Dekany}, {Drake}, {Fremling}, {Healy}, {Karambelkar}, {Kasliwal}, {Kong}, {Kulkarni}, {Masci}, {Post}, {Purdum}, {Rich}, \& {Wold}}]{2024A&A...684L..18B}
{Brennan}, S.~J., {Sollerman}, J., {Irani}, I., {et~al.} 2024, \aap, 684, L18, \dodoi{10.1051/0004-6361/202449350}

\bibitem[{{Brunetti} {et~al.}(2004){Brunetti}, {Sanchez del Rio}, {Golosio}, {Simionovici}, \& {Somogyi}}]{2004AcSpB..59.1725B}
{Brunetti}, A., {Sanchez del Rio}, M., {Golosio}, B., {Simionovici}, A., \& {Somogyi}, A. 2004, Spectrochimica Acta - Part B: Atomic Spectroscopy, 59, 1725, \dodoi{10.1016/j.sab.2004.03.014}

\bibitem[{{Chandra} {et~al.}(2012){Chandra}, {Chevalier}, {Chugai}, {Fransson}, {Irwin}, {Soderberg}, {Chakraborti}, \& {Immler}}]{2012ApJ...755..110C}
{Chandra}, P., {Chevalier}, R.~A., {Chugai}, N., {et~al.} 2012, \apj, 755, 110, \dodoi{10.1088/0004-637X/755/2/110}

\bibitem[{{Chevalier}(1982)}]{1982ApJ...258..790C}
{Chevalier}, R.~A. 1982, ApJ, 258, 790, \dodoi{10.1086/160126}

\bibitem[{{Chevalier} \& {Fransson}(2003)}]{2003LNP...598..171C}
{Chevalier}, R.~A., \& {Fransson}, C. 2003, in Supernovae and Gamma-Ray Bursters, ed. K.~{Weiler}, Vol. 598, 171--194, \dodoi{10.1007/3-540-45863-8_10}

\bibitem[{{Chevalier} \& {Fransson}(2006)}]{2006ApJ...651..381C}
---. 2006, \apj, 651, 381, \dodoi{10.1086/507606}

\bibitem[{{Chevalier} \& {Fransson}(2017)}]{2017hsn..book..875C}
---. 2017, in Handbook of Supernovae, ed. A.~W. {Alsabti} \& P.~{Murdin}, 875, \dodoi{10.1007/978-3-319-21846-5_34}

\bibitem[{{Chevalier} \& {Irwin}(2012)}]{2012ApJ...747L..17C}
{Chevalier}, R.~A., \& {Irwin}, C.~M. 2012, \apjl, 747, L17, \dodoi{10.1088/2041-8205/747/1/L17}

\bibitem[{{Chugai}(2009)}]{2009MNRAS.400..866C}
{Chugai}, N.~N. 2009, \mnras, 400, 866, \dodoi{10.1111/j.1365-2966.2009.15506.x}

\bibitem[{{Cox} \& {Tucker}(1969)}]{1969ApJ...157.1157C}
{Cox}, D.~P., \& {Tucker}, W.~H. 1969, \apj, 157, 1157, \dodoi{10.1086/150144}

\bibitem[{{Davis} {et~al.}(2023){Davis}, {Taggart}, {Tinyanont}, {Foley}, {Villar}, {Izzo}, {Angus}, {Bustamante-Rosell}, {Coulter}, {Earl}, {Farias}, {Hjorth}, {Huber}, {Jones}, {Kelly}, {Kilpatrick}, {Langeroodi}, {Miao}, {Pellegrino}, {Ramirez-Ruiz}, {Ransome}, {Rest}, {Siebert}, {Terreran}, {Thornton}, {Yadavalli}, {Zeimann}, {Auchettl}, {Bom}, {Brink}, {Burke}, {Camacho-Neves}, {Chambers}, {de Boer}, {DeMarchi}, {Filippenko}, {Galbany}, {Gall}, {Gao}, {Herpich}, {Howell}, {Jacobson-Galan}, {Jha}, {Kanaan}, {Khetan}, {Kwok}, {Lai}, {Larison}, {Lin}, {Loertscher}, {Magnier}, {McCully}, {McGill}, {Newsome}, {Padilla Gonzalez}, {Pan}, {Rest}, {Rho}, {Ribeiro}, {Santos}, {Schoenell}, {Sharief}, {Smith}, {Wainscoat}, {Wang}, {Zenati}, \& {Zheng}}]{2023MNRAS.523.2530D}
{Davis}, K.~W., {Taggart}, K., {Tinyanont}, S., {et~al.} 2023, \mnras, 523, 2530, \dodoi{10.1093/mnras/stad1433}

\bibitem[{{Dessart} \& {Hillier}(2010)}]{2010MNRAS.405.2141D}
{Dessart}, L., \& {Hillier}, D.~J. 2010, \mnras, 405, 2141, \dodoi{10.1111/j.1365-2966.2010.16611.x}

\bibitem[{{Dessart} {et~al.}(2022){Dessart}, {Hillier}, \& {Kuncarayakti}}]{2022A&A...658A.130D}
{Dessart}, L., {Hillier}, D.~J., \& {Kuncarayakti}, H. 2022, \aap, 658, A130, \dodoi{10.1051/0004-6361/202142436}

\bibitem[{{Drury}(1983)}]{1983RPPh...46..973D}
{Drury}, L.~O. 1983, Reports on Progress in Physics, 46, 973, \dodoi{10.1088/0034-4885/46/8/002}

\bibitem[{{Fermi}(1949)}]{1949PhRv...75.1169F}
{Fermi}, E. 1949, Physical Review, 75, 1169, \dodoi{10.1103/PhysRev.75.1169}

\bibitem[{{Filippenko}(1997)}]{1997ARA&A..35..309F}
{Filippenko}, A.~V. 1997, \araa, 35, 309, \dodoi{10.1146/annurev.astro.35.1.309}

\bibitem[{{Foley} {et~al.}(2007){Foley}, {Smith}, {Ganeshalingam}, {Li}, {Chornock}, \& {Filippenko}}]{2007ApJ...657L.105F}
{Foley}, R.~J., {Smith}, N., {Ganeshalingam}, M., {et~al.} 2007, \apjl, 657, L105, \dodoi{10.1086/513145}

\bibitem[{{Fransson}(1982)}]{1982A&A...111..140F}
{Fransson}, C. 1982, \aap, 111, 140

\bibitem[{{Fransson} \& {Bj{\"o}rnsson}(1998)}]{1998ApJ...509..861F}
{Fransson}, C., \& {Bj{\"o}rnsson}, C.-I. 1998, \apj, 509, 861, \dodoi{10.1086/306531}

\bibitem[{{Fransson} {et~al.}(1996){Fransson}, {Lundqvist}, \& {Chevalier}}]{1996ApJ...461..993F}
{Fransson}, C., {Lundqvist}, P., \& {Chevalier}, R.~A. 1996, \apj, 461, 993, \dodoi{10.1086/177119}

\bibitem[{{Fransson} {et~al.}(2014){Fransson}, {Ergon}, {Challis}, {Chevalier}, {France}, {Kirshner}, {Marion}, {Milisavljevic}, {Smith}, {Bufano}, {Friedman}, {Kangas}, {Larsson}, {Mattila}, {Benetti}, {Chornock}, {Czekala}, {Soderberg}, \& {Sollerman}}]{2014ApJ...797..118F}
{Fransson}, C., {Ergon}, M., {Challis}, P.~J., {et~al.} 2014, \apj, 797, 118, \dodoi{10.1088/0004-637X/797/2/118}

\bibitem[{{Gal-Yam} {et~al.}(2021){Gal-Yam}, {Yaron}, {Pastorello}, {Taubenberger}, {Fraser}, \& {Perley}}]{2021TNSAN..76....1G}
{Gal-Yam}, A., {Yaron}, O., {Pastorello}, A., {et~al.} 2021, Transient Name Server AstroNote, 76, 1

\bibitem[{{Gal-Yam} {et~al.}(2022){Gal-Yam}, {Bruch}, {Schulze}, {Yang}, {Perley}, {Irani}, {Sollerman}, {Kool}, {Soumagnac}, {Yaron}, {Strotjohann}, {Zimmerman}, {Barbarino}, {Kulkarni}, {Kasliwal}, {De}, {Yao}, {Fremling}, {Yan}, {Ofek}, {Fransson}, {Filippenko}, {Zheng}, {Brink}, {Copperwheat}, {Foley}, {Brown}, {Siebert}, {Leloudas}, {Cabrera-Lavers}, {Garcia-Alvarez}, {Marante-Barreto}, {Frederick}, {Hung}, {Wheeler}, {Vink{\'o}}, {Thomas}, {Graham}, {Duev}, {Drake}, {Dekany}, {Bellm}, {Rusholme}, {Shupe}, {Andreoni}, {Sharma}, {Riddle}, {van Roestel}, \& {Knezevic}}]{2022Natur.601..201G}
{Gal-Yam}, A., {Bruch}, R., {Schulze}, S., {et~al.} 2022, \nat, 601, 201, \dodoi{10.1038/s41586-021-04155-1}

\bibitem[{{Gangopadhyay} {et~al.}(2020){Gangopadhyay}, {Misra}, {Hiramatsu}, {Wang}, {Hosseinzadeh}, {Wang}, {Valenti}, {Zhang}, {Howell}, {Arcavi}, {Anupama}, {Burke}, {Dastidar}, {Itagaki}, {Kumar}, {Kumar}, {Li}, {McCully}, {Mo}, {Pandey}, {Pellegrino}, {Sai}, {Sahu}, {Sanwal}, {Singh}, {Singh}, {Zhang}, {Zhang}, \& {Zhang}}]{2020ApJ...889..170G}
{Gangopadhyay}, A., {Misra}, K., {Hiramatsu}, D., {et~al.} 2020, \apj, 889, 170, \dodoi{10.3847/1538-4357/ab6328}

\bibitem[{{Hatchett} {et~al.}(1976){Hatchett}, {Buff}, \& {McCray}}]{1976ApJ...206..847H}
{Hatchett}, S., {Buff}, J., \& {McCray}, R. 1976, \apj, 206, 847, \dodoi{10.1086/154448}

\bibitem[{{Hosseinzadeh} {et~al.}(2017){Hosseinzadeh}, {Arcavi}, {Valenti}, {McCully}, {Howell}, {Johansson}, {Sollerman}, {Pastorello}, {Benetti}, {Cao}, {Cenko}, {Clubb}, {Corsi}, {Duggan}, {Elias-Rosa}, {Filippenko}, {Fox}, {Fremling}, {Horesh}, {Karamehmetoglu}, {Kasliwal}, {Marion}, {Ofek}, {Sand}, {Taddia}, {Zheng}, {Fraser}, {Gal-Yam}, {Inserra}, {Laher}, {Masci}, {Rebbapragada}, {Smartt}, {Smith}, {Sullivan}, {Surace}, \& {Wo{\'z}niak}}]{2017ApJ...836..158H}
{Hosseinzadeh}, G., {Arcavi}, I., {Valenti}, S., {et~al.} 2017, \apj, 836, 158, \dodoi{10.3847/1538-4357/836/2/158}

\bibitem[{{Immler} {et~al.}(2008){Immler}, {Modjaz}, {Landsman}, {Bufano}, {Brown}, {Milne}, {Dessart}, {Holland}, {Koss}, {Pooley}, {Kirshner}, {Filippenko}, {Panagia}, {Chevalier}, {Mazzali}, {Gehrels}, {Petre}, {Burrows}, {Nousek}, {Roming}, {Pian}, {Soderberg}, \& {Greiner}}]{2008ApJ...674L..85I}
{Immler}, S., {Modjaz}, M., {Landsman}, W., {et~al.} 2008, \apjl, 674, L85, \dodoi{10.1086/529373}

\bibitem[{{Itoh}(1977)}]{1977PASJ...29..813I}
{Itoh}, H. 1977, \pasj, 29, 813

\bibitem[{{Kallman} \& {McCray}(1982)}]{1982ApJS...50..263K}
{Kallman}, T.~R., \& {McCray}, R. 1982, \apjs, 50, 263, \dodoi{10.1086/190828}

\bibitem[{{Longair}(2011)}]{2011hea..book.....L}
{Longair}, M.~S. 2011, {High Energy Astrophysics}

\bibitem[{{Maeda}(2012)}]{2012ApJ...758...81M}
{Maeda}, K. 2012, \apj, 758, 81, \dodoi{10.1088/0004-637X/758/2/81}

\bibitem[{{Maeda}(2013)}]{2013ApJ...762...14M}
---. 2013, \apj, 762, 14, \dodoi{10.1088/0004-637X/762/1/14}

\bibitem[{{Maeda} {et~al.}(2014){Maeda}, {Katsuda}, {Bamba}, {Terada}, \& {Fukazawa}}]{2014ApJ...785...95M}
{Maeda}, K., {Katsuda}, S., {Bamba}, A., {Terada}, Y., \& {Fukazawa}, Y. 2014, \apj, 785, 95, \dodoi{10.1088/0004-637X/785/2/95}

\bibitem[{{Maeda} \& {Moriya}(2022)}]{2022ApJ...927...25M}
{Maeda}, K., \& {Moriya}, T.~J. 2022, \apj, 927, 25, \dodoi{10.3847/1538-4357/ac4672}

\bibitem[{{Matsuoka} {et~al.}(2019){Matsuoka}, {Maeda}, {Lee}, \& {Yasuda}}]{2019ApJ...885...41M}
{Matsuoka}, T., {Maeda}, K., {Lee}, S.-H., \& {Yasuda}, H. 2019, \apj, 885, 41, \dodoi{10.3847/1538-4357/ab4421}

\bibitem[{{Mattila} {et~al.}(2008){Mattila}, {Meikle}, {Lundqvist}, {Pastorello}, {Kotak}, {Eldridge}, {Smartt}, {Adamson}, {Gerardy}, {Rizzi}, {Stephens}, \& {van Dyk}}]{2008MNRAS.389..141M}
{Mattila}, S., {Meikle}, W.~P.~S., {Lundqvist}, P., {et~al.} 2008, \mnras, 389, 141, \dodoi{10.1111/j.1365-2966.2008.13516.x}

\bibitem[{{Moriya} \& {Maeda}(2016)}]{2016ApJ...824..100M}
{Moriya}, T.~J., \& {Maeda}, K. 2016, \apj, 824, 100, \dodoi{10.3847/0004-637X/824/2/100}

\bibitem[{{Moriya} {et~al.}(2013){Moriya}, {Maeda}, {Taddia}, {Sollerman}, {Blinnikov}, \& {Sorokina}}]{2013MNRAS.435.1520M}
{Moriya}, T.~J., {Maeda}, K., {Taddia}, F., {et~al.} 2013, \mnras, 435, 1520, \dodoi{10.1093/mnras/stt1392}

\bibitem[{{Moriya} {et~al.}(2014){Moriya}, {Maeda}, {Taddia}, {Sollerman}, {Blinnikov}, \& {Sorokina}}]{2014MNRAS.439.2917M}
---. 2014, \mnras, 439, 2917, \dodoi{10.1093/mnras/stu163}

\bibitem[{{Morozova} {et~al.}(2015){Morozova}, {Piro}, {Renzo}, {Ott}, {Clausen}, {Couch}, {Ellis}, \& {Roberts}}]{2015ApJ...814...63M}
{Morozova}, V., {Piro}, A.~L., {Renzo}, M., {et~al.} 2015, \apj, 814, 63, \dodoi{10.1088/0004-637X/814/1/63}

\bibitem[{{Murase} {et~al.}(2019){Murase}, {Franckowiak}, {Maeda}, {Margutti}, \& {Beacom}}]{2019ApJ...874...80M}
{Murase}, K., {Franckowiak}, A., {Maeda}, K., {Margutti}, R., \& {Beacom}, J.~F. 2019, \apj, 874, 80, \dodoi{10.3847/1538-4357/ab0422}

\bibitem[{{Nagao} {et~al.}(2023){Nagao}, {Kuncarayakti}, {Maeda}, {Moore}, {Pastorello}, {Mattila}, {Uno}, {Smartt}, {Sim}, {Ferrari}, {Tomasella}, {Anderson}, {Chen}, {Galbany}, {Gao}, {Gromadzki}, {Guti{\'e}rrez}, {Inserra}, {Kankare}, {Magnier}, {M{\"u}ller-Bravo}, {Reguitti}, \& {Young}}]{2023A&A...673A..27N}
{Nagao}, T., {Kuncarayakti}, H., {Maeda}, K., {et~al.} 2023, \aap, 673, A27, \dodoi{10.1051/0004-6361/202346084}

\bibitem[{{Nomoto} \& {Hashimoto}(1988)}]{1988PhR...163...13N}
{Nomoto}, K., \& {Hashimoto}, M. 1988, Physics Reports, 163, 13, \dodoi{https://doi.org/10.1016/0370-1573(88)90032-4}

\bibitem[{{Nozawa} {et~al.}(2008){Nozawa}, {Kozasa}, {Tominaga}, {Sakon}, {Tanaka}, {Suzuki}, {Nomoto}, {Maeda}, {Umeda}, {Limongi}, \& {Onaka}}]{2008ApJ...684.1343N}
{Nozawa}, T., {Kozasa}, T., {Tominaga}, N., {et~al.} 2008, \apj, 684, 1343, \dodoi{10.1086/589961}

\bibitem[{{Ofek} {et~al.}(2007){Ofek}, {Cameron}, {Kasliwal}, {Gal-Yam}, {Rau}, {Kulkarni}, {Frail}, {Chandra}, {Cenko}, {Soderberg}, \& {Immler}}]{2007ApJ...659L..13O}
{Ofek}, E.~O., {Cameron}, P.~B., {Kasliwal}, M.~M., {et~al.} 2007, \apjl, 659, L13, \dodoi{10.1086/516749}

\bibitem[{{Ofek} {et~al.}(2014){Ofek}, {Zoglauer}, {Boggs}, {Barri{\'e}re}, {Reynolds}, {Fryer}, {Harrison}, {Cenko}, {Kulkarni}, {Gal-Yam}, {Arcavi}, {Bellm}, {Bloom}, {Christensen}, {Craig}, {Even}, {Filippenko}, {Grefenstette}, {Hailey}, {Laher}, {Madsen}, {Nakar}, {Nugent}, {Stern}, {Sullivan}, {Surace}, \& {Zhang}}]{2014ApJ...781...42O}
{Ofek}, E.~O., {Zoglauer}, A., {Boggs}, S.~E., {et~al.} 2014, \apj, 781, 42, \dodoi{10.1088/0004-637X/781/1/42}

\bibitem[{{Pastorello} {et~al.}(2007){Pastorello}, {Smartt}, {Mattila}, {Eldridge}, {Young}, {Itagaki}, {Yamaoka}, {Navasardyan}, {Valenti}, {Patat}, {Agnoletto}, {Augusteijn}, {Benetti}, {Cappellaro}, {Boles}, {Bonnet-Bidaud}, {Botticella}, {Bufano}, {Cao}, {Deng}, {Dennefeld}, {Elias-Rosa}, {Harutyunyan}, {Keenan}, {Iijima}, {Lorenzi}, {Mazzali}, {Meng}, {Nakano}, {Nielsen}, {Smoker}, {Stanishev}, {Turatto}, {Xu}, \& {Zampieri}}]{2007Natur.447..829P}
{Pastorello}, A., {Smartt}, S.~J., {Mattila}, S., {et~al.} 2007, \nat, 447, 829, \dodoi{10.1038/nature05825}

\bibitem[{{Pastorello} {et~al.}(2008){Pastorello}, {Mattila}, {Zampieri}, {Della Valle}, {Smartt}, {Valenti}, {Agnoletto}, {Benetti}, {Benn}, {Branch}, {Cappellaro}, {Dennefeld}, {Eldridge}, {Gal-Yam}, {Harutyunyan}, {Hunter}, {Kjeldsen}, {Lipkin}, {Mazzali}, {Milne}, {Navasardyan}, {Ofek}, {Pian}, {Shemmer}, {Spiro}, {Stathakis}, {Taubenberger}, {Turatto}, \& {Yamaoka}}]{2008MNRAS.389..113P}
{Pastorello}, A., {Mattila}, S., {Zampieri}, L., {et~al.} 2008, \mnras, 389, 113, \dodoi{10.1111/j.1365-2966.2008.13602.x}

\bibitem[{{Pellegrino} {et~al.}(2022{\natexlab{a}}){Pellegrino}, {Howell}, {Terreran}, {Arcavi}, {Bostroem}, {Brown}, {Burke}, {Dong}, {Gilkis}, {Hiramatsu}, {Hosseinzadeh}, {McCully}, {Modjaz}, {Newsome}, {Gonzalez}, {Pritchard}, {Sand}, {Valenti}, \& {Williamson}}]{2022ApJ...938...73P}
{Pellegrino}, C., {Howell}, D.~A., {Terreran}, G., {et~al.} 2022{\natexlab{a}}, \apj, 938, 73, \dodoi{10.3847/1538-4357/ac8ff6}

\bibitem[{{Pellegrino} {et~al.}(2022{\natexlab{b}}){Pellegrino}, {Howell}, {Vink{\'o}}, {Gangopadhyay}, {Xiang}, {Arcavi}, {Brown}, {Burke}, {Hiramatsu}, {Hosseinzadeh}, {Li}, {McCully}, {Misra}, {Newsome}, {Gonzalez}, {Pritchard}, {Valenti}, {Wang}, \& {Zhang}}]{2022ApJ...926..125P}
{Pellegrino}, C., {Howell}, D.~A., {Vink{\'o}}, J., {et~al.} 2022{\natexlab{b}}, \apj, 926, 125, \dodoi{10.3847/1538-4357/ac3e63}

\bibitem[{{Pellegrino} {et~al.}(2024){Pellegrino}, {Modjaz}, {Takei}, {Tsuna}, {Newsome}, {Pritchard}, {Baer-Way}, {Bostroem}, {Chandra}, {Charalampopoulos}, {Dong}, {Farah}, {Howell}, {McCully}, {Mohamed}, {Padilla Gonzalez}, \& {Terreran}}]{2024arXiv240718291P}
{Pellegrino}, C., {Modjaz}, M., {Takei}, Y., {et~al.} 2024, arXiv e-prints, arXiv:2407.18291, \dodoi{10.48550/arXiv.2407.18291}

\bibitem[{{Perley} {et~al.}(2022){Perley}, {Sollerman}, {Schulze}, {Yao}, {Fremling}, {Gal-Yam}, {Ho}, {Yang}, {Kool}, {Irani}, {Yan}, {Andreoni}, {Baade}, {Bellm}, {Brink}, {Chen}, {Cikota}, {Coughlin}, {Dahiwale}, {Dekany}, {Duev}, {Filippenko}, {Hoeflich}, {Kasliwal}, {Kulkarni}, {Lunnan}, {Masci}, {Maund}, {Medford}, {Riddle}, {Rosnet}, {Shupe}, {Strotjohann}, {Tzanidakis}, \& {Zheng}}]{2022ApJ...927..180P}
{Perley}, D.~A., {Sollerman}, J., {Schulze}, S., {et~al.} 2022, \apj, 927, 180, \dodoi{10.3847/1538-4357/ac478e}

\bibitem[{{Pursiainen} {et~al.}(2023){Pursiainen}, {Leloudas}, {Schulze}, {Charalampopoulos}, {Angus}, {Anderson}, {Bauer}, {Chen}, {Galbany}, {Gromadzki}, {Guti{\'e}rrez}, {Inserra}, {Lyman}, {M{\"u}ller-Bravo}, {Nicholl}, {Smartt}, {Tartaglia}, {Wiseman}, \& {Young}}]{2023ApJ...959L..10P}
{Pursiainen}, M., {Leloudas}, G., {Schulze}, S., {et~al.} 2023, \apjl, 959, L10, \dodoi{10.3847/2041-8213/ad103d}

\bibitem[{{Rybicki} \& {Lightman}(1979)}]{1979rpa..book.....R}
{Rybicki}, G.~B., \& {Lightman}, A.~P. 1979, {Radiative processes in astrophysics}

\bibitem[{{Sakon} {et~al.}(2009){Sakon}, {Onaka}, {Wada}, {Ohyama}, {Kaneda}, {Ishihara}, {Tanab{\'e}}, {Minezaki}, {Yoshii}, {Tominaga}, {Nomoto}, {Nozawa}, {Kozasa}, {Tanaka}, {Suzuki}, {Umeda}, {Ohyabu}, {Usui}, {Matsuhara}, {Nakagawa}, \& {Murakami}}]{2009ApJ...692..546S}
{Sakon}, I., {Onaka}, T., {Wada}, T., {et~al.} 2009, \apj, 692, 546, \dodoi{10.1088/0004-637X/692/1/546}

\bibitem[{{Schoonjans} {et~al.}(2011){Schoonjans}, {Brunetti}, {Golosio}, {Sanchez del Rio}, {Sol{\'e}}, {Ferrero}, \& {Vincze}}]{2011AcSpB..66..776S}
{Schoonjans}, T., {Brunetti}, A., {Golosio}, B., {et~al.} 2011, Spectrochimica Acta - Part B: Atomic Spectroscopy, 66, 776, \dodoi{10.1016/j.sab.2011.09.011}

\bibitem[{{Shivvers} {et~al.}(2016){Shivvers}, {Zheng}, {Mauerhan}, {Kleiser}, {Van Dyk}, {Silverman}, {Graham}, {Kelly}, {Filippenko}, \& {Kumar}}]{2016MNRAS.461.3057S}
{Shivvers}, I., {Zheng}, W.~K., {Mauerhan}, J., {et~al.} 2016, \mnras, 461, 3057, \dodoi{10.1093/mnras/stw1528}

\bibitem[{{Shivvers} {et~al.}(2017){Shivvers}, {Zheng}, {Van Dyk}, {Mauerhan}, {Filippenko}, {Smith}, {Foley}, {Mazzali}, {Kamble}, {Kilpatrick}, {Margutti}, {Yuk}, {Graham}, {Kelly}, {Andrews}, {Matheson}, {Wood-Vasey}, {Ponder}, {Brown}, {Chevalier}, {Milisavljevic}, {Drout}, {Parrent}, {Soderberg}, {Ashall}, {Piascik}, \& {Prentice}}]{2017MNRAS.471.4381S}
{Shivvers}, I., {Zheng}, W., {Van Dyk}, S.~D., {et~al.} 2017, \mnras, 471, 4381, \dodoi{10.1093/mnras/stx1885}

\bibitem[{{Smith}(2014)}]{2014ARA&A..52..487S}
{Smith}, N. 2014, \araa, 52, 487, \dodoi{10.1146/annurev-astro-081913-040025}

\bibitem[{{Smith}(2017)}]{2017hsn..book..403S}
---. 2017, in Handbook of Supernovae, ed. A.~W. {Alsabti} \& P.~{Murdin}, 403, \dodoi{10.1007/978-3-319-21846-5_38}

\bibitem[{{Smith} {et~al.}(2007){Smith}, {Li}, {Foley}, {Wheeler}, {Pooley}, {Chornock}, {Filippenko}, {Silverman}, {Quimby}, {Bloom}, \& {Hansen}}]{2007ApJ...666.1116S}
{Smith}, N., {Li}, W., {Foley}, R.~J., {et~al.} 2007, \apj, 666, 1116, \dodoi{10.1086/519949}

\bibitem[{{Stritzinger} {et~al.}(2012){Stritzinger}, {Taddia}, {Fransson}, {Fox}, {Morrell}, {Phillips}, {Sollerman}, {Anderson}, {Boldt}, {Brown}, {Campillay}, {Castellon}, {Contreras}, {Folatelli}, {Habergham}, {Hamuy}, {Hjorth}, {James}, {Krzeminski}, {Mattila}, {Persson}, \& {Roth}}]{2012ApJ...756..173S}
{Stritzinger}, M., {Taddia}, F., {Fransson}, C., {et~al.} 2012, \apj, 756, 173, \dodoi{10.1088/0004-637X/756/2/173}

\bibitem[{{Sun} {et~al.}(2020){Sun}, {Maund}, {Hirai}, {Crowther}, \& {Podsiadlowski}}]{2020MNRAS.491.6000S}
{Sun}, N.-C., {Maund}, J.~R., {Hirai}, R., {Crowther}, P.~A., \& {Podsiadlowski}, P. 2020, \mnras, 491, 6000, \dodoi{10.1093/mnras/stz3431}

\bibitem[{{Tarter} {et~al.}(1969){Tarter}, {Tucker}, \& {Salpeter}}]{1969ApJ...156..943T}
{Tarter}, C.~B., {Tucker}, W.~H., \& {Salpeter}, E.~E. 1969, \apj, 156, 943, \dodoi{10.1086/150026}

\bibitem[{{Tominaga} {et~al.}(2008){Tominaga}, {Limongi}, {Suzuki}, {Tanaka}, {Nomoto}, {Maeda}, {Chieffi}, {Tornambe}, {Minezaki}, {Yoshii}, {Sakon}, {Wada}, {Ohyama}, {Tanab{\'e}}, {Kaneda}, {Onaka}, {Nozawa}, {Kozasa}, {Kawabata}, {Anupama}, {Sahu}, {Gurugubelli}, {Prabhu}, \& {Deng}}]{2008ApJ...687.1208T}
{Tominaga}, N., {Limongi}, M., {Suzuki}, T., {et~al.} 2008, \apj, 687, 1208, \dodoi{10.1086/591782}

\bibitem[{{Tsuna} {et~al.}(2021){Tsuna}, {Kashiyama}, \& {Shigeyama}}]{2021ApJ...914...64T}
{Tsuna}, D., {Kashiyama}, K., \& {Shigeyama}, T. 2021, \apj, 914, 64, \dodoi{10.3847/1538-4357/abfaf8}

\end{thebibliography}
\bibliographystyle{aasjournal}

\end{document}